\documentclass[twocolumn]{aastex62}
\usepackage{color,soul}
\usepackage{amsmath}
\usepackage{comment}

\defcitealias{Kennicutt1998}{K98}
\defcitealias{RC3}{RC3}
\defcitealias{PT05}{PT05}
\defcitealias{delosReyes2019}{Paper I}
\defcitealias{DraineLi2007}{DL07}

\newcommand{\hi}{H\textsc{i}}

\newcommand{\xco}{$X(\textrm{CO})$}

\newcommand{\msun}{M$_\odot$}

\graphicspath{{./}{figures/}}

\submitjournal{ApJ}

\shorttitle{Paper II: The Integrated Star Formation Law for Starbursts}
\shortauthors{Kennicutt et al.}

\begin{document}

\title{Revisiting the Integrated Star Formation Law. II.  Starbursts and the 
    Combined Global Schmidt Law}  

\correspondingauthor{Robert C. Kennicutt Jr.}

\author{Robert C. Kennicutt Jr.}
\affiliation{Department of Astronomy and Steward Observatory, University of Arizona \\
Tucson, AZ 85721, USA}
\affiliation{Department of Physics and Astronomy, Texas A\&M University \\
College Station, TX 77843, USA}
\affiliation{Institute of Astronomy, University of Cambridge \\
Madingley Road, Cambridge CB3 0HA, UK}

\author[0000-0002-4739-046X]{Mithi A. C. de los Reyes}
\affiliation{Department of Astronomy, California Institute of Technology \\
1200 E. California Blvd., MC 249-17 \\
Pasadena, CA 91125, USA }
\affiliation{Institute of Astronomy, University of Cambridge \\
Madingley Road, Cambridge CB3 0HA, UK}

\begin{abstract}
We compile observations of molecular gas contents and infrared-based star formation rates (SFRs) for 112
circumnuclear star forming regions, in order to re-investigate the form of the disk-averaged Schmidt surface density star formation law in starbursts.  We then combine these results with total gas and SFR surface densities for 153 nearby non-starbursting disk galaxies from de los Reyes \& Kennicutt (2019), to investigate the properties of the combined star formation law, following Kennicutt (1998; K98).  We confirm that the combined Schmidt law can be fitted with a single power law with slope $n = 1.5\pm0.05$ (including fitting method uncertainties), somewhat steeper than the value $n = 1.4\pm0.15$ found by K98.  Fitting separate power laws to the non-starbursting and starburst galaxies, however, produces very different slopes ($n = 1.34\pm0.07$ and $0.98\pm0.07$, respectively), with a pronounced offset in the zeropoint ($\sim$0.6\,dex) of the starburst relation to higher SFR surface densities.  This offset is seen even when a common conversion factor between CO intensity and molecular hydrogen surface density is applied, and is confirmed when disk surface densities of interstellar dust are used as proxies for gas measurements. Tests for possible systematic biases in the starburst data fail to uncover any spurious sources for such a large offset.  We tentatively conclude that the global Schmidt law in galaxies, at least as it is conventionally measured, is bimodal or possibly  multi-modal.  Possible causes may include changes in the small-scale structure of the molecular ISM or the stellar initial mass function.  A single $n \sim 1.5$ power law still remains as a credible approximation or ``recipe" for analytical or numerical models of galaxy formation and evolution.
\end{abstract}


\section{Introduction} 
\label{sec:intro}

The empirical relationship between the rate of star formation and density of cold gas in galaxies, usually expressed in terms of surface densities ($\Sigma$), is a primary ingredient in models and simulations of galaxy formation and evolution.
The physical processes which trigger and regulate the star formation rate (SFR) on galactic scales are far too complex to accurately model on a purely theoretical basis, so simulations need to rely on sub-grid empirical prescriptions, which have been applied with considerable success \citep[e.g.,][]{Vogelsberger2014,Schaye2015}.  

Since the first proposals by \citet{Schmidt1959} and \citet{Schmidt1963}, the most commonly applied parameterization of the star formation law has been a power law:
\begin{equation}
\Sigma_{\textrm{SFR}} = A(\Sigma_{\textrm{gas}})^{n}
\label{eq:SchmidtLaw}
\end{equation}

\citet{Kennicutt1998} (hereafter \citetalias{Kennicutt1998}) demonstrated that the disk-averaged surface densities of star formation and gas in nearby galaxies are well fitted by a Schmidt power law with slope $n=1.4\pm 0.15$.
This power-law relation extended over five orders of magnitude in gas density (seven in SFR surface density), from the extended disks of normal star-forming spirals to the more luminous and compact circumnuclear disks found in luminous and ultraluminous starburst galaxies.
The same data were fit nearly as well by a linear correlation between SFR surface density and the ratio of gas surface density to mean dynamical time (orbit time) of the systems, a relation suggested previously by \citet{Silk97} and \citet{Elmegreen97}, which we will refer to hereafter as the Silk-Elmegreen law.

The relative tightness and apparent universality of these empirical star formation laws, at least on global scales, led to their widespread application observationally and theoretically.
The relations calibrated in \citetalias{Kennicutt1998} were based on relatively sparse datasets: 61 normal star-forming disks and only 36 infrared-selected starburst galaxies.  
Subsequent surveys of \hi{}, CO, and SFRs in normal and starburst galaxies (as traced in the ultraviolet and infrared) have greatly expanded the available samples of systems with spatially-resolved measurements, and made it possible to conduct a critical re-examination of the form, dispersion, and continuity of the global star formation law.  

This is the second of two papers in which we aim to re-examine the results from \citetalias{Kennicutt1998}.
In \citet{delosReyes2019} (hereafter \citetalias{delosReyes2019}) we used updated measurements of \hi{}, CO, and ultraviolet (UV) and infrared (IR) based SFRs for 165 normal (i.e., non-starbursting galaxies on the main blue sequence) spiral and 161 dwarf galaxies to re-investigate the form of the star formation law in the lower-density regime.
We found that the SFR and total gas surface densities are well represented by an $n = 1.41 \pm 0.07$ power law, consistent with the relationship found by \citetalias{Kennicutt1998} for the entire normal to starburst galaxy ensemble.
On the other hand we found that the dependence of SFR surface density on disk-averaged molecular gas density is much shallower, and consistent with the behavior of the spatially-resolved relation within similar types of galaxies \citep[e.g.,][]{Leroy2008, Bigiel2011}.  
The new dataset also shows evidence for a sharp falloff or threshold in the SFR at very low gas surface densities, a result most clearly seen in low surface brightness spiral and dwarf irregular galaxies \citep[also see][]{Wyder2009,Shi2014}, and again echoing thresholds observed previously in spatially-resolved studies \citep[e.g.,][]{Kennicutt1989, Martin2001, Bigiel2008}.  
A multi-parametric analysis revealed a significant second-order correlation between the residuals in the Schmidt density law with the surface density of stars in disks, similar to the result found earlier by \citet{Shi2011} and \citet{Shi2018}.

In this second paper we focus on the star formation law in the higher-density regime found in the circumnuclear disks of starburst galaxies, and in the dense molecular circumnuclear disks of many strongly barred galaxies.
Our new sample of 114 galaxies more than triples the number in \citetalias{Kennicutt1998}.  The new sample also takes advantage of surveys over the past 20 years with instruments on the IRAM, ALMA, Spitzer, and Herschel observatories to compile a more consistent set of measurements of the gas masses, radii, surface densities, and SFRs than was possible previously.

We then combine the data from \citetalias{delosReyes2019} and this paper to reanalyze the continuity of the star formation law over the full range of surface densities in the full galaxy sample, and address specific issues which have arisen since the publication of \citetalias{Kennicutt1998}.
Among these are the reports of substantially shallower slopes than the $n \sim 1.4$ value found in \citetalias{Kennicutt1998} \citep[e.g.,][]{Genzel2010, Liu2015}, and suggestions of separate or bimodal star
formation laws separating normal star-forming disks from dense luminous starbursts \citep[e.g.,][]{Daddi2010,Genzel2010}.  The degree of bimodality seen in these studies is strongly dependent on the values of the CO-to-H$_2$ conversion factor [\xco{}] applied to the different classes of galaxies.  When a systematic shift is applied in \xco{} the Schmidt laws separate by a roughly corresponding amount, while adoption of a density-dependent \xco{} correction can produce a more highly non-linear relation \citep{Narayanan2012}.  A final paper in this series (Paper III, in preparation) will extend our analysis to a larger and more diverse set of galaxies, using measurements of dust masses and surface densities as proxies for the corresponding \hi{} and CO measurements.

Some aspects of our analysis closely parallel a previously published  investigation by \citet{Liu2015}, who used radio continuum (synchrotron) based SFRs to also study the behavior of the global star formation law for a larger sample than in \citetalias{Kennicutt1998}, and we will compare the two sets of results throughout this paper.  

The remainder of this paper is organized as follows.  In \S\ref{sec:data}\ we present our new compilation of molecular gas masses, SFRs, disk radii, and respective surface densities, as derived from CO and infrared flux measurements in the literature.
In \S\ref{sec:results}\ we examine the form of the Schmidt power-law relationships and Silk-Elmegreen relationships, both for the starbursts considered alone and the full merged galaxy sample, and re-examine the evidence for a monotonic vs. a multi-modal law.  
In \S\ref{sec:dustSchmidt}\ we use surface densities of interstellar dust in a subset of the galaxies to perform a consistency check on the results based on total gas surface densities, and demonstrate that dust measurements offer a potentially powerful new tool for investigating the form of the global star formation law (Paper III in this series, in preparation).  
We compare our results with those from previous
studies in \S\ref{sec:literature}.
In \S\ref{sec:discussion} we discuss the physical consequences of the new results, and offer recommendations on the usability and limitations of these empirical star formation prescriptions for observational and theoretical applications.
In \S\ref{sec:conclusion} we briefly summarize our results.

\section{Data} 
\label{sec:data}
The basic approach of this analysis was to probe the high-density regime of the star formation law by selecting a sample of galaxies with compact circumnuclear starbursts, as described in \citetalias{Kennicutt1998}.

\subsection{Galaxy Sample}
\label{sec:sample}

The core of our sample is a set of dusty starburst galaxies identified by their total IR (TIR) luminosities either as luminous infrared galaxies (LIRGs) with $\log(\textrm{TIR}) \ge 10^{11}$ L$_\odot$, or as ultraluminous infrared galaxies (ULIRGs) with $\log(\textrm{TIR}) \ge 10^{12}$ L$_\odot$.
We adopt the conventional definition of TIR luminosity to be that integrated over the wavelength range $8 - 1000\,\mu$m \citep{Sanders1996}.  For this study we only considered those LIRGs and ULIRGs with centrally concentrated star formation (generally $r = 0.3 - 3$ kpc), as determined from high-resolution CO or infrared imaging.

Despite a considerable growth in observations of these galaxies over the past 20 years, the scope of this study is still limited by the availability of both CO observations and high-resolution infrared observations, which are needed to measure the relevant gas masses, dust luminosities, region sizes, and thence the corresponding surface densities of young stars and gas.  Consequently our sample is not drawn from any single survey, but rather is a compendium of all of the galaxies with such measurements available in the literature.  Nearly all of the LIRGs and ULIRGs in our sample, however, are contained in the comprehensive Spitzer Great Observatories All-Sky LIRG Survey (GOALS) \citep{Armus2009}, and we adopt the basic catalog data from that survey (e.g., positions, distances, infrared luminosities) here. 

Once a candidate set of infrared galaxies with the necessary CO and IR data was compiled, a number of other tests were applied to ensure that the galaxies in this sample represented compact circumnuclear disks dominated by recent star formation.  LIRGs and ULIRGs can be powered by a combination of massive star formation bursts and luminous dust-obscured active galactic nuclei (AGN), and it was important to remove from our sample galaxies with large AGN contributions.  This was accomplished by using the ionization of the mid-infrared fine structure lines to separate stellar versus accretion-dominated heating sources \citep[e.g.,][]{Petric2011,Alonso-Herrero2012,Inami2013}.  A number of AGN tracers are available including [Ne\textsc{v}]14.3\,$\mu$m, [O\textsc{iv}25.9\,$\mu$m] in comparison to lower-ionization species such as [Ne\textsc{ii}]12.8\,$\mu$m, analyzed alone or in combination with the equivalent widths of the aromatic PAH features in the mid-infrared (see \citet{Petric2011}).  For example galaxies with [Ne\textsc{v}]/[Ne\textsc{ii}] $\ge$ 0.2 were excluded, which corresponds to a limit of $\sim$15\%\ of dust heating from the AGN \citep{Petric2011}. Such measurements were available for most of the galaxies in the sample; when in doubt we erred in the direction of excluding a galaxy where there was a possibility of an significant ($>$15\%) AGN contamination.  This resulted in the rejection of approximately 15\%\ of our original sample, but this number does not include widely-known examples of radio or X-ray-loud AGNs/LIRGs which were excluded from our study from the beginning.   \citep{Petric2011} report that across the entire GOALS sample (selected independently from the presence of AGNs) only 12\%\ of the dust luminosity is from AGNs. After our additional screening for AGNs the contamination should be considerably less.

We also used Herschel Space Observatory images of the galaxies \citep{Chu2017} to exclude a handful of LIRGs in which the star formation is distributed over an extended disk as opposed to the circumnuclear regions, or systems with multiple components where the infrared and CO measurements could not be unambiguously matched to each other.  After applying both of these tests, a total of 89 star-formation dominated LIRGs and ULIRGs with the necessary CO, infrared, and spatial information were included in our sample (out of an original list of approximately 115 candidates).

We supplement our sample of LIRGs and ULIRGs with observations of the dense circumnuclear disks of 23 otherwise normal spiral galaxies.  Selection was based on the availability of high-resolution CO interferometer maps, and most are barred galaxies with dense central molecular disks \citep[e.g.,][]{Sakamoto1999} or well-observed starbursting circumnuclear disks which fall below the luminosity threshold for LIRGs (e.g, NGC 253, M82).  The central disks in the barred galaxies are related to circumnuclear ``hotspot" galaxies originally identified by \citet{Morgan1958} and \citet{Sersic1965}.  The central molecular disks are generally characterized by average surface densities $\Sigma_{H_2} > 100~\mathrm{M}_{\odot}~\mathrm{pc}^{-2}$ and much higher than a central extrapolation of the outer disk density profile.  Unlike the LIRGs and ULIRGs in this sample, the star formation contained in these local circumnuclear disks rarely dominates the total amount of star formation in the respective galaxies, but as will be seen later their surface densities lie in a similar regime to those of the more luminous starbursts.   Many these objects are the dense centers of galaxies with integrated measurements reported in \citetalias{delosReyes2019}, and included in our control sample of normal spiral disks.  

Throughout this paper we will be comparing the star formation, gas, and dust surface densities in the circumnuclear starbursts with those of for a sample of normal, non-starbursting galaxies from \citetalias{delosReyes2019}.  Complete data on the global properties of their galaxies, their HI and H$_2$ gas contents, star formation rates, and surface densities will be found in that paper. The \citetalias{delosReyes2019} sample consists of 307 spiral and dwarf galaxies.  We have extracted from that paper 153
spiral galaxies with M$_B < -18$ and stellar masses above $10^9$\,M$_{\odot}$.  The intent of this selection is to ensure that the comparison galaxies span similar ranges of masses and metal abundances to the starburst galaxies (or their progenitors).  It is worth noting that \citet{delosReyes2019} tested for and found no significant second-order dependence of the Schmidt law relation on galaxy mass (see their Figure 9), so the conclusions of this paper should not be sensitive to the mass or luminosity cutoffs applied to the normal galaxy sample.

\subsection{CO Fluxes and Gas Masses}
\label{sec:gas}

In contrast to the extended disks of most normal spiral galaxies, where the masses of atomic and molecular gas are comparable, observations of the dense circumnuclear regions of dusty starburst galaxies show that the disks are overwhelmingly molecular dominated \citep[e.g.,][]{Sanders1996}.  
Therefore following \citetalias{Kennicutt1998} we will use the molecular gas surface density as a proxy for the total gas density.  Any systematic uncertainty introduced by this approximation is negligible in comparison to uncertainties in the value of the \xco{}{} conversion factor itself.
We note that this approximation only applies to the central starburst regions themselves; at larger radii in the same galaxies the atomic mass can easily approach or sometimes exceed the total galaxy mass in molecular material.  

Our estimates of the molecular gas masses and or surface density distributions are mostly based on measurements of the CO(1--0) rotational line, or on the CO(2--1) line when it was the lowest level transition available, assuming a CO(2--1)/CO(1--0) ratio of 0.8 \citep{Leroy2009}.  
Whenever possible we used published interferometric observations of the gas from the Owens Valley Radio Observatory, Plateau de Bure interferometer, or the Atacama Large Millimeter Array (ALMA) to directly map the column density distribution of the molecular gas.  A total of 60 starbursts are in this ``direct'' subsample.

Comparisons of the spatial distribution of the gas with the infrared-emitting dust or the star-forming regions traced in Pa\,$\alpha$ generally show excellent agreement among measurements of the spatial extent of the circumnuclear disk (\S\ref{sec:diameters}). This enabled us to estimate mean molecular surface densities for an additional set of galaxies with integrated single-dish CO measurements, when information on the size of the central disk was available from other wavelengths and it could be ascertained from these images that the circumnuclear region accounted for most of the integrated molecular emission within the single-dish aperture (12--50\arcsec\ HPBW).  We refer to the 52 galaxies with surface densities derived in this way as the ``indirect'' CO subsample.  

For some of the galaxies with only single-dish measurements the beam sizes are considerably larger than the sizes of the circumnuclear starbursts, and it is likely that the CO fluxes include emission from the surrounding regions.  As detailed in \S\ref{sec:IR} below, data from the Spitzer and Herschel observatories were used to estimate the fraction of emission external to the defined circumnuclear regions ($f_{\mathrm{ext}}$).  We then applied the same corrections to the single-dish CO fluxes whenever the beam sizes were substantially larger than the diameters of the circumnuclear disks, on the assumption that the CO and infrared emissions roughly trace each other.  This is an approximation, but in view of the magnitude of the corrections any deviations from the assumptions should not have a significant effect on our results.

Key to the interpretation of the star formation law for starburst galaxies is the choice of conversion factor \xco{} between CO line intensity and molecular hydrogen surface density.  Assumptions adopted in the literature differ widely, as reviewed by \citet{Bolatto2013} and \citet{Tacconi2020}.  In K98 a constant solar neighborhood value of \xco{} was adopted purely for the sake of simplicity, and a number of recent studies have suggested that such a choice is realistic (\citet{Tacconi2020} and references therein.  On the other hand, studies of multiple CO rotational transitions in LIRGs and ULIRGs suggest a higher CO emissivity in those objects (i.e., significantly lower values of \xco{}), and applying significantly lower conversions tends to produce a bimodal Schmidt law (e.g., \citet{Daddi2010}, \citet{Genzel2010}). 
For this study we shall investigate the sensitivity of the star formation law on the presumed application of \xco{} by showing results for the full range of possibilities, ranging from a constant local value for all galaxies:

\begin{equation}
X(\textrm{CO}) =  2.0 \times 10^{20}~\textrm{cm}^{-2}~(\textrm{K~km~s}^{-1})^{-1},
\label{eq:XCO_MW}
\end{equation}

\noindent
to a systematically lower value value for starbursts, and a density-dependent prescription recommended by \citet{Bolatto2013}; see \S3.3).  For the comparison sample of normal star-forming spirals we uniformly adopt the conversion in eq\,[2] above.  This neglects a small possible dependence of \xco{} on galaxy mass and mean metal abundance, but we justify the approximation on the fact that most of the spirals lie in the regime where \xco{} is constant or very weakly dependent on metallicity \citep{Bolatto2013}.  We already demonstrated in \citet{delosReyes2019} that application of a metallicity-dependent prescription to that galaxy sample does not significantly alter the derived Schmidt law.


Following \citetalias{delosReyes2019} we will present surface densities for hydrogen alone, but later when deriving gas depletion times we include a correction factor of 1.36 to account for helium and other molecules.

\subsection{Infrared Fluxes and Star Formation Rates}
\label{sec:IR}

The typical dust attenuation in compact LIRG and ULIRG sources is very high. For example for the GOALS sample of local LIRGs, the median contribution of (unobscured) far-ultraviolet (FUV) emission to the total SFR is only 2.8\%\ (average 4\%), and it is only 18\%\ even in the least dusty system in the sample \citep{Howell2010}.  These high attenuations are consistent with the very high surface densities of the central disks as measured in CO, assuming a typical dust-to-gas ratio.  As a result we will base our SFR measurements for the LIRGs and ULIRGs solely on their total infrared (TIR: $8 - 1000$\,$\mu$m)  dust luminosities.  

In most published studies of the star formation law for LIRGs and ULIRGs, including \citetalias{Kennicutt1998}, the relevant TIR luminosities were taken from the IRAS survey, which used apertures of order 4\farcm5 in diameter, more than an order of magnitude larger than the radii of even the largest starburst regions.  This approximation is often justified, as the circumnuclear regions usually dominate the integrated light of their respective galaxies, especially the most luminous systems.  The publication of spatially-resolved measurements of most of these galaxies from the Spitzer Space Telescope \citep[e.g.,][]{Diaz-Santos2010} and the Herschel Space Observatory \citep[e.g.,][]{Lutz2016} make it possible to test the validity of this approximation and, when appropriate, to apply approximate aperture corrections from IRAS (or use the Herschel-based fluxes directly).  

\citet{Diaz-Santos2010} published FWHM sizes for the infrared-emitting regions at 13.2\,$\mu$m along with a fraction of extended emission (FEE), while \citet{Lutz2016} published effective radii at 70, 100, and 160\,$\mu$m. We used these measurements along with the GOALS Herschel image atlas of \citet{Chu2017} to check the disk sizes estimated in other ways (\S\ref{sec:diameters}) and, whenever necessary, to correct the IRAS total fluxes for emission outside of our circumnuclear regions.  These measurements were also helpful for resolving multiple sources within the IRAS beams, which are common in these predominantly interaction and merger-driven systems.  The Herschel measurements at 100\,$\mu$m were applied because \citet{Galametz2013} has shown that the monochromatic luminosity at that wavelength correlates most strongly with total infrared luminosity over a wide range of galaxy types. We refer to our extended emission corrections as $f_{\mathrm{ext}}$ to distinguish them from the FEE values in \citet{Diaz-Santos2010}, which are defined differently and sometimes correspond to different region sizes.  Since we deliberately chose disk diameters which include most if not all of the circumnuclear emission (and removed any LIRGs from the sample with substantial extended disk star formation) these corrections were generally small; 30 of the 91 LIRGs/ULIRGs required corrections of $0.25 < f_{\mathrm{ext}} \le 0.5$, 11 have corrections of $\le0.25$, and for 47 no correction is needed. The highest $f_{\mathrm{ext}}$ corrections often apply to double or multiple merging systems, where the higher resolution Herschel or Spitzer observations make it possible to measure individual component fluxes \citep[e.g.,][]{Chu2017}.

Star formation rates were estimated from the corrected TIR luminosities using the calibration published in 
\citet{KennicuttEvans2012}:

\begin{equation}
\log[\mathrm{SFR}~(\mathrm{M}_{\odot}~\mathrm{yr}^{-1})] = \log[L(\mathrm{TIR})_{\mathrm{corr}}~(\mathrm{erg}~\mathrm{s}^{-1})] - 43.41
\end{equation}
The dust-heating
stellar population is assumed to follow a \citet{Kroupa1993} IMF (mass limits 0.1 -- 100\,\msun),
with continuous star formation over
a period of 100 Myr (mean age 50 Myr).  This conversion gives SFRs that are 14\%\ lower than the
frequently applied calibration of \citet{KennicuttARAA98}, which used older stellar models and a 
Salpeter IMF.  This equation implicitly assumes that the dust has absorbed all of the starlight from 
the burst and that the dust heating is dominated by the newly formed stars (i.e., no significant
contributions from an older stellar population or AGN).  
As was the case with the gas masses, the derived SFRs are sensitive to the 
assumed star formation history and IMF, and the effects of varying these assumptions will be
discussed in \S\ref{sec:discussion}.

Although many of the circumnuclear disks in the local sample show similarly high dust attenuations, IRAS fluxes are usually inappropriate for estimating the circumnuclear SFRs, either because the centers only contribute a small part of the global infrared emission, or because the dust emission does not reprocess all of the young starlight.  Instead we derived SFRs using a combination of other traces, including UV, H$\alpha$, Pa$\alpha$, mid-infrared, or radio continuum fluxes as available.  In most cases these measurements were taken from the original papers providing the CO data, with updated calibrations from \citet{KennicuttEvans2012} applied.  In a few galaxies with bright centrally-dominated TIR emission (e.g., M82, NGC 253) the IRAS fluxes could be applied directly, with aperture corrections as described above. We discuss the overall uncertainties in these measurements below in \S\ref{sec:uncertainties}.

\subsection{Diameters of the Star Formation Regions and Surface Densities}
\label{sec:diameters}

By definition circumnuclear starburst regions have radial extents that are a small fraction of 
their host galaxy sizes. In order to provide a physically meaningful measure of the surface
densities in the star-forming regions, these need to be based on the distributions of 
gas, dust, or young stars themselves.  In \citetalias{Kennicutt1998} these were mainly taken from 
millimeter interferometer maps, on the assumption that nearly all of the integrated TIR emission of
the galaxy was emitted from the same region.  

Over the past twenty years more spatially resolved measurements of the CO as well as the infrared continuum and Pa\,$\alpha$ line emission have become available, and these allow us to compare the sizes of the molecular and star-forming disks directly.  We compiled data for 24 starbursts (mostly LIRGs and ULIRGs) for which there were interferometric measurements in CO available as well as sizes for the IR-emitting regions (see \S\ref{sec:IR} above) and/or from Pa\,$\alpha$ emission
\citep{Alonso-Herrero2006,Tateuchi2015}.  For this sample the median ratio of the diameter of the dust-emitting or ionized gas disk to the diameter of the CO-emitting disk was 0.99, confirming the validity of the usual assumption that the respective disk sizes are similar, at least on a statistical basis.  The dispersion in diameter
ratios is significant, however, $\pm$0.23 dex for this sample.  Much of this dispersion arises from uncertainties in the measurements themselves (infrared disk sizes at different wavelengths show a typical scatter of $\pm$20-25\%), but some appears to reflect the fact that the dust and CO emission do not trace each other exactly.  Whenever there was a significant difference in diameter between wavelengths we opted for the larger of the choices, to ensure that the apertures adopted included most of the circumnuclear emission at all wavelengths.

For the galaxies lacking interferometer CO maps we used the measured sizes of the circumnuclear disks in the far-infrared, mid-infrared, Pa$\alpha$, H$\alpha$, or near-infrared (as measured with HST) as proxies, in that order of preference. In \S\ref{sec:literature} we will explore the consequences of changing these assumptions.

Surface densities of gas and SFR were defined simply as the gas masses
and SFRs divided by $\pi\,a^2$, where $a$ is the semi-major axis of the star-forming disk 
(this implicitly applies the inclination correction in the thin-disk approximation, which was
adopted for simplicity).

\subsection{Dust Masses and Surface Densities}
\label{sec:dust}

In \S\ref{sec:dustSchmidt} we explore the use of dust surface densities of the star-forming disks as proxies for CO-derived measurements of the gas densities.  We used dust masses measured for the GOALS sample of LIRGs and ULIRGs by \citet{Shangguan2019}.  These were measured by fitting a \citet{DraineLi2007} model to the far-infrared and submillimeter SEDs of the galaxies, and rescaled to an updated zeropoint \citep{Draine2014}, as discussed more fully in \S\ref{sec:dustSchmidt}.  The dust masses were based on integrated infrared photometry of the GOALS galaxies, and as a result any aperture corrections applied to the dust luminosities and SFRs (\S\ref{sec:IR}) were applied identically to the dust masses. Dust surface densities were calculated by scaling the dust masses to the diameters of the disks discussed above.

\subsection{Other Parameters}
\label{sec:otherparams}

One of the main results from \citetalias{Kennicutt1998} was the demonstration that the global star formation law is well
represented by a ``Silk-Elmegreen'' relation in which the SFR surface density scales linearly with
the ratio of gas surface density to dynamical time (taken here to be the orbit time at the radius
of the star-forming disk).  In order to re-investigate this result, rotation velocities were compiled
for the subset of 40 starburst regions where the CO maps had sufficient spatial resolution and showed
evidence for rotationally-dominated motions.  These were corrected for inclination and combined with
the radii to measure the dynamical times.  The subsample includes both LIRGs and local starbursts,
with a significantly higher proportional representation of the latter, because of the prevalence of
disturbed velocity fields in many of the LIRGs and ULIRGs (many are young merger remnants).

Readers may note that our data tabulation does not include measurements of stellar masses for the starbursts.  Stellar masses (and the specific star formation rates which can be derived from the ratio of SFRs and stellar masses) are important parameters for characterizing any star-forming system, and they formed a key element of our own analysis of the star formation law for normal disks in \citetalias{delosReyes2019}.  In the case of the infrared-luminous starbursts, however, we know of no way to accurately measure stellar masses for the central starburst regions which are relevant here.  The very high surface densities in these regions (of order $10^2 - 10^4$ M$_{\odot}$\,pc$^{-2}$, or column densities of order $10^{22} - 10^{24}$\,cm$^{-2}$), correspond to line-of-sight dust attenuations of order 10--1000 mag in the visible, and several magnitudes even in the near-infrared.  This combined with the uncertain age distributions of stars in these often-disturbed systems makes a quantitative stellar mass determination in the central regions problematic.  As mentioned above we do estimate central dynamical masses for those systems where reliable kinematic data are available, but given the large molecular masses and mass fractions in the regions, the dynamical information only allows us to impose very crude statistical limits on the stellar masses at best.

Estimating stellar masses for the host galaxies of the starbursts is less
problematic, and such masses were derived for the LIRGs and ULIRGs in this 
sample by \citet{Shangguan2019}.  The masses range between $10^{10.3} - 
10^{11.5}$ \msun, with an average near $10^{10.8}$ \msun.  This is similar to the range for the comparison sample of spiral galaxies.

\subsection{Data Summary and Uncertainties}
\label{sec:uncertainties}

Table~\ref{tab:starburst} presents a summary of the general properties for the circumnuclear starburst sample. 

Column 1: running index number.

Column 2: galaxy name preferred either by the NASA/IPAC Extragalactic Database (NED)\footnote{The NASA/IPAC Extragalactic Database (NED) is operated by the Jet Propulsion Laboratory, California Institute of Technology, under contract with the National Aeronautics and Space Administration.}.

Columns 3 and 4 (hidden in print version): J2000 right ascension and declination as reported in NED.

Column 5: luminosity distance, taken from GOALS \citep{Armus2009} for the LIRGs and ULIRGs, or from the sources referenced for the local barred galaxies (assuming H$_0$ = 70\,km\,s$^{-1}$\,Mpc$^{-1}$).  Since we are mainly interested in analyzing (distance-independent) surface densities, the precise distances are not important.  

Column 6: adopted diameter, as described in \S\ref{sec:diameters}.

Column 7: radio diameter, as described in \S\ref{sec:literature}.

Column 8: total infrared luminosity, taken from GOALS \citep{Armus2009} whenever possible.

Column 9: star formation rate, either computed from the total infrared luminosity or from multi-wavelength observations described in \S\ref{sec:IR}.

Column 10: fraction of IR emission outside of the defined circumnuclear region  $f_{\mathrm{ext}}$, as described in \S\ref{sec:IR}. These corrections have already been applied to the data in this table (columns 8, 9, 10, and 12).

Column 11: H$_{2}$ mass, as described in \S\ref{sec:gas}.  

Column 12: references for H$_{2}$ mass. The full references are given in Appendix~\ref{appendix:gasrefs}.

Column 13: dust mass, as described in \S\ref{sec:dust} and \S\ref{sec:dustSchmidt}.

Column 14: flag denoting if interferometric maps were available (1 if true, 0 if ``indirect'' single-dish measurements were used).

\bigskip

Table~\ref{tab:derived} lists the derived quantities used in our analysis for the starburst sample, including surface densities.

Columns 1 and 2: running index number and galaxy name.

Column 3: H$_{2}$ gas surface density.

Column 4: star formation rate surface density.

Column 5: dust surface density.

Column 6: inclination-corrected rotational velocity at the outer radius of the circumnuclear disk, as derived from CO measurements.

Column 7: dynamical time, defined as the orbit time at the outer radius of the circumnuclear disk (\S\ref{sec:otherparams}).

\begin{deluxetable*}
{llhhlllllllclc}
\tablecolumns{14} 
\tablecaption{Basic properties of starburst sample. \label{tab:starburst}} 
\tablehead{ 
\colhead{N} & \colhead{ID} & \nocolhead{RA} & \nocolhead{Dec} & \colhead{Dist} & \colhead{$D$} & \colhead{$D_{\mathrm{radio}}$}  & \colhead{$\log L_{\mathrm{IR}}$} &
\colhead{$\log\mathrm{SFR}$\tablenotemark{a}} & \colhead{$f_{\mathrm{ext}}$} & \colhead{$\log M_{\mathrm{H}_{2}}$} & \colhead{H$_{2}$ ref.} & \colhead{$\log M_{\mathrm{dust}}$} & \colhead{Mapped} \\
\colhead{} & \colhead{} & \nocolhead{(J2000)} & \nocolhead{(J2000)} & \colhead{(Mpc)} & \colhead{(\arcsec)} & \colhead{(\arcsec)} & \colhead{([$L_{\odot}$])} &
\colhead{([$M_{\odot}~\mathrm{y}^{-1}$])} & \colhead{} & \colhead{([$M_{\odot}$])} & \colhead{} & \colhead{([$M_{\odot}$])} &\colhead{}
}
\startdata
\multicolumn{14}{c}{Local circumnuclear disks}\\[0.5em]
\tableline
1 & NGC 253 & 00h47m33.12s & -25d17m17.6s & 3.4 & 44.0 & \ldots & 9.97 & 0.64 & 0.42 & 7.78 & 5 & \ldots & 1 \\
2 & NGC 470 & 01h19m44.85s & +03d24m35.8s & 30.5 & 14.0 & \ldots & \ldots & 0.48 & 0.0 & 8.55 & 8 & \ldots & 1 \\
3 & NGC 1097 & 02h46m19.05s & -30d16m29.6s & 16.0 & 24.0 & \ldots & \ldots & 0.53 & 0.0 & 9.14 & 14 & \ldots & 1 \\
4 & NGC 1365 & 03h33m36.37s & -36d08m25.4s & 18.0 & 22.0 & \ldots & \ldots & 0.82 & 0.0 & 9.42 & 15 & \ldots & 1 \\
5 & IC 342 & 03h46m48.50s & +68d05m46.9s & 3.0 & 66.0 & \ldots & \ldots & -0.47 & 0.0 & 8.04 & 16 & \ldots & 1 \\
\tableline
\multicolumn{14}{c}{LIRGs}\\[0.5em]
\tableline
24 & UGC 6 & 00h03m09.62s & +21d57m36.6s & 92.0 & 8.0 & \ldots & 10.91 & 1.10 & 0.0 & 9.38 & 1 & \ldots & 1 \\
25 & IRAS F00057+4021 & 00h08m20.47s & +40d37m56.7s & 165.0 & 2.4 & \ldots & 11.6 & 1.79 & 0.0 & 10.19 & 2 & \ldots & 1 \\
26 & NGC 23  & 00h09m53.41s & +25d55m25.6s & 65.0 & 7.1 & 8.0 & 10.82 & 1.31 & 0.5 & 9.26 & 3,4 & 7.51 & 0 \\
27 & NGC 34 & 00h11m06.55s & -12d06m26.3s & 83.0 & 6.0 & 0.4 & 11.41 & 1.60 & 0.0 & 10.04 & 1 & 7.62 & 1 \\
28 & MCG+12-02-001 & 00h54m03.61s & +73d05m11.8s & 70.0 & 6.0 & 3.2 & 11.35 & 1.69 & 0.3 & 9.79 & 3 & 7.68 & 0 \\
\enddata
\tablenotetext{a}{For most of the local circumnuclear disks, SFRs are derived from multi-wavelength observations as described in the text, rather than from IR luminosities. We therefore list SFRs and not IR luminosities for these galaxies.}
\tablecomments{Only a portion of Table~\ref{tab:starburst} is shown here; it is published in its entirety (including coordinates) in the machine-readable format online.}
\tablereferences{See Appendix~\ref{appendix:gasrefs}}
\end{deluxetable*}

\begin{deluxetable*}
{lllDlll}
\tablecolumns{7} 
\tablecaption{Derived properties of starburst sample. \label{tab:derived}} 
\tablehead{ 
\colhead{N} & \colhead{ID} &  \colhead{$\log \Sigma_{\mathrm{H}_{2}}$}  & \multicolumn2c{$\log \Sigma_{\mathrm{SFR}}$} & \colhead{$\log \Sigma_{\mathrm{dust}}$} & \colhead{$v_{\mathrm{rot}}$} & \colhead{$t_{\mathrm{dyn}}$} \\
\colhead{} & \colhead{} & \colhead{([$M_{\odot}~\mathrm{pc}^{-2}$])} & \multicolumn2c{($[M_{\odot}~\mathrm{yr}^{-1} \mathrm{kpc}^{-2}]$)} &
\colhead{($[M_{\odot}~\mathrm{pc}^{-2}]$)} & \colhead{($\mathrm{km}~\mathrm{s}^{-1}$)} & \colhead{($10^{8}~\mathrm{y}$)}
}
\decimals
\startdata
\tableline
\multicolumn{7}{c}{Local circumnuclear disks}\\[0.5em]
\tableline
1 & NGC 253 & 2.16 & 0.79 & \ldots & 135 & 0.17 \\
2 & NGC 470 & 2.02 & -0.05 & \ldots & 131 & 0.49 \\
3 & NGC 1097 & 2.71 & 0.1 & \ldots & 200 & 0.29 \\
4 & NGC 1365 & 2.96 & 0.36 & \ldots & 150 & 0.39 \\
5 & IC 342 & 2.18 & -0.33 & \ldots & \ldots & \ldots \\
\tableline
\multicolumn{7}{c}{LIRGs}\\[0.5em]
\tableline
24 & UGC 6 & 2.38 & 0.1 & \ldots & 307 & 0.36 \\
25 & IRAS F00057+4021 & 3.73 & 1.33 & \ldots & 250 & 0.24 \\
26 & NGC 23  & 2.66 & 0.41 & 0.91 & \ldots & \ldots \\
27 & NGC 34 & 3.38 & 0.94 & 0.96 & 250 & 0.3 \\
28 & MCG+12-02-001 & 3.27 & 1.02 & 1.16 & \ldots & \ldots \\
\enddata
\tablecomments{Only a portion of Table~\ref{tab:derived} is shown here; it is published in its entirety in the machine-readable format online.}
\end{deluxetable*}

\bigskip

As mentioned in previous sections, the uncertainties in these measurements arise from a 
combination of observational and systematic sources, and we discuss their approximate magnitudes here.  
Note that the sensitivity of the star formation law to systematic
uncertainties will be addressed at more length in \S\ref{sec:literature} and \S\ref{sec:discussion}, after the main results are presented.

For molecular gas masses, the potential sources of observational uncertainty include extended flux missed by
the interferometer for the mapped regions, and potential contributions of gas lying outside of
the circumnuclear disk for the single-dish measurements.  We restricted the mapped sample to 
data where missing interferometer flux was not expected to be a problem, and checked against 
single-dish measurements whenever possible.  Regions with suspect fluxes were not included in the
sample.  Likewise single-dish measurements were only included where it could be determined 
{\it a priori} that circumnuclear emission was the dominant contributor to the observed flux.
We estimate that those factors taken together introduce uncertainties of the order of $\pm$0.1\,dex ($\pm$25\%) into the total fluxes. 
This is comparable to the differences between independent measurements of the same objects.
The largest systematic uncertainty is in the scaling factor \xco{} used to convert molecular
line intensities to mass estimates.  

Most of LIRGs and ULIRGs are strong infrared emitters, so the precision of the fluxes themselves
was not a significant source of uncertainty.  As mentioned earlier the assumption of 
nearly complete dust attenuation for the LIRG sample has been confirmed by \citet{Howell2010}.  Instead the main observational uncertainties arise
from the use of integrated fluxes from IRAS, combined with uncertainties in any aperture corrections
applied to those fluxes (\S\ref{sec:IR}).  For a majority of the starbursts the circumnuclear disks are 
compact and well isolated from the IR emission from the rest of the galaxy.    In the cases where
significant aperture corrections were required ($\sim$35\%\ of the sample), uncertainties in the corrections
could introduce uncertainties as large as $\pm$20\%\ in the resulting SFRs.  

As with the molecular
gas masses the larger sources of uncertainty are systematic in nature.  The conversion of TIR
luminosity to SFR is sensitive to the stellar IMF, the assumed opacity of the dust in the region,
and the star formation history of the dust heating population, all of which will vary somewhat from system to system.  We adopt a Kroupa IMF for 
consistency with \citetalias{delosReyes2019}, and touch on the possible impacts of deviations from this assumption in the discussion later (\S\ref{sec:discussion}).  
For the local starburst sample we relied on custom measurements
of the SFR, often from the original papers (converted to consistent IMF and SFR recipes).
Comparison of multiple measurements generally shows agreement to $\pm$30\%\ or better.

A significant source of systematic error in the SFRs is introduced by variations in the ages of the starbursts.  Evolutionary
synthesis models \citep[e.g.,][]{Leitherer99} show that the bolometric luminosity of a 
continuously star-forming population increases by $\sim$30\%\ between ages of 10 to 100\,Myr,
so variations in SFRs for a fixed luminosity at the level of $\pm$10--20\%\ should be expected.
Dust heating by older stellar populations and/or AGNs should be minimal (our selection criteria should have eliminated galaxies with large contributions from evolved stars), but contributions of up to 10--20\%\ cannot be ruled out in all cases.  

Uncertainties in the diameters of the circumnuclear regions will propagate to the measured surface densities, 
independent of uncertainties in the respective total fluxes.  To the extent that the dust and gas emission are spatially correlated, however, uncertainties in diameters will affect both sets of surface densities identically, effectively moving the points along the trajectory of a linear star formation law.  These problems should not be very significant for this paper but may affect some previously published studies, as discussed in \S\ref{sec:literature}.

Combining the sources of observational uncertainty we are confident in assigning conservative
errors of no more than $\pm$0.2 dex to both the molecular hydrogen and SFR surface densities.
This is acceptable when considering that the surface densities of the starburst galaxies span $>$2 dex, and that the dynamic range of the combined spiral galaxy and starburst sample is 4.5--6 dex in gas and SFR surface densities, respectively.  
Potential systematic errors will be discussed later.

\begin{figure}
    \centering
    \epsscale{1.15}
    \plotone{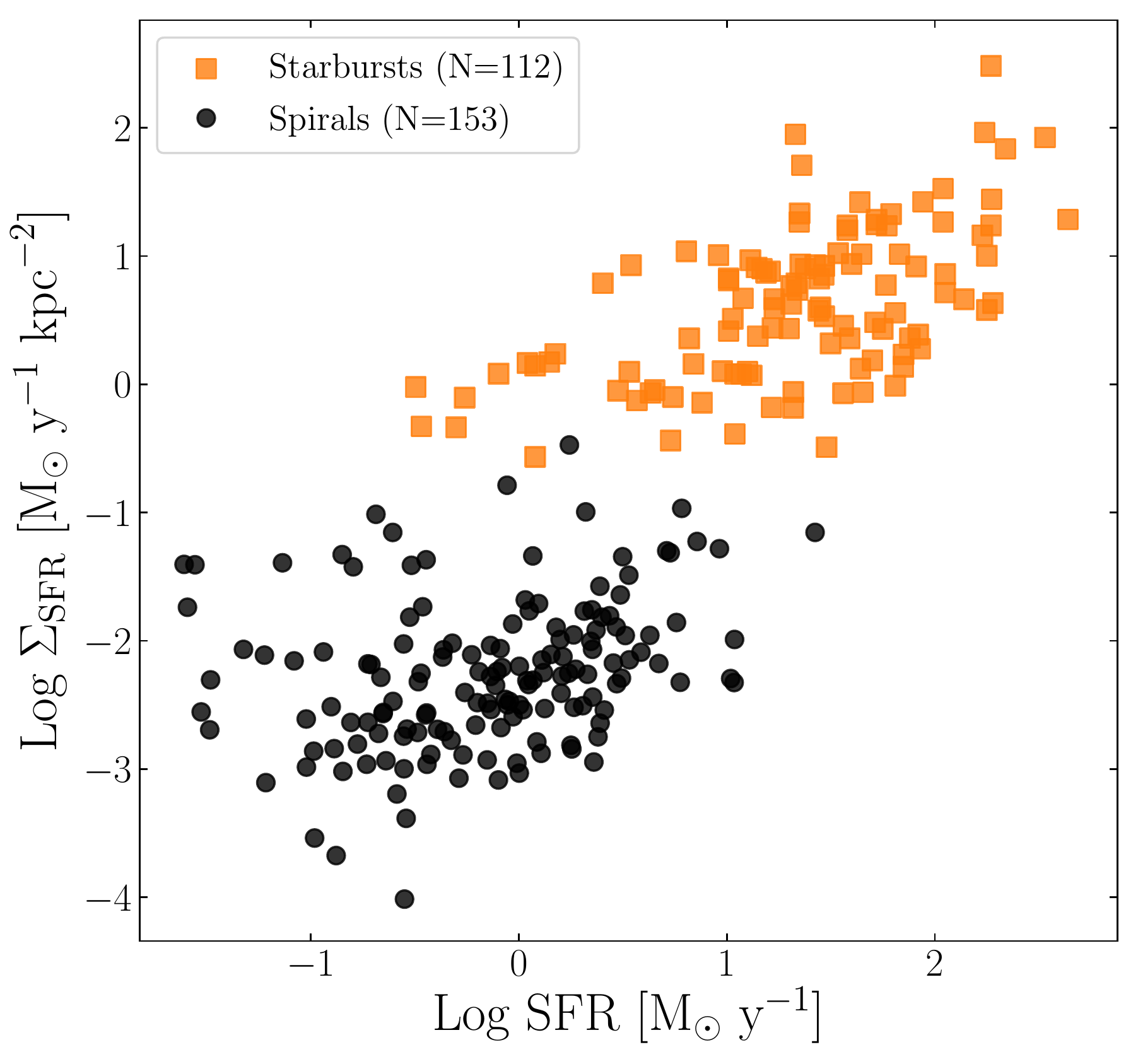}
    \caption{Comparison of absolute SFRs and SFR surface densities for the starburst galaxies (orange squares) and the comparison sample of normal spiral disks (black circles) from \citetalias{delosReyes2019}.}
    \label{fig:sampleprops}
\end{figure}

Figure~\ref{fig:sampleprops} compares the distributions of absolute star formation rates (SFRs) and SFR surface densities
of the starburst sample and the comparison sample of non-starbursting
(blue sequence) spiral galaxy disks, the latter taken from \citetalias{delosReyes2019}.  The LIRG and ULIRG samples are physically
distinct from the normal spirals both in terms of their absolute SFRs (the luminosity threshold of a
LIRG corresponds to a SFR of $\sim$16\,\msun\,yr$^{-1}$, higher than the equilibrium SFR in all but
the most massive undisturbed spiral galaxies), and in terms of the surface density of star formation.  Both of these properties are defining characteristics of starburst galaxies, so the offsets in both axes of Figure~\ref{fig:sampleprops} should not be surprising.
The local circumnuclear starbursts tend to overlap in surface density with the LIRGs/ULIRGs
but with lower absolute SFRs.

\section{Results}
\label{sec:results}

We begin by examining the form of the star formation law for the starburst sample on its own, and then investigate the combined relation for non-starbursting disk galaxies and the starbursts.

\subsection{The Schmidt Law for Circumnuclear Starbursts}

\begin{figure*}
    \centering
    \epsscale{1.1}
    \plottwo{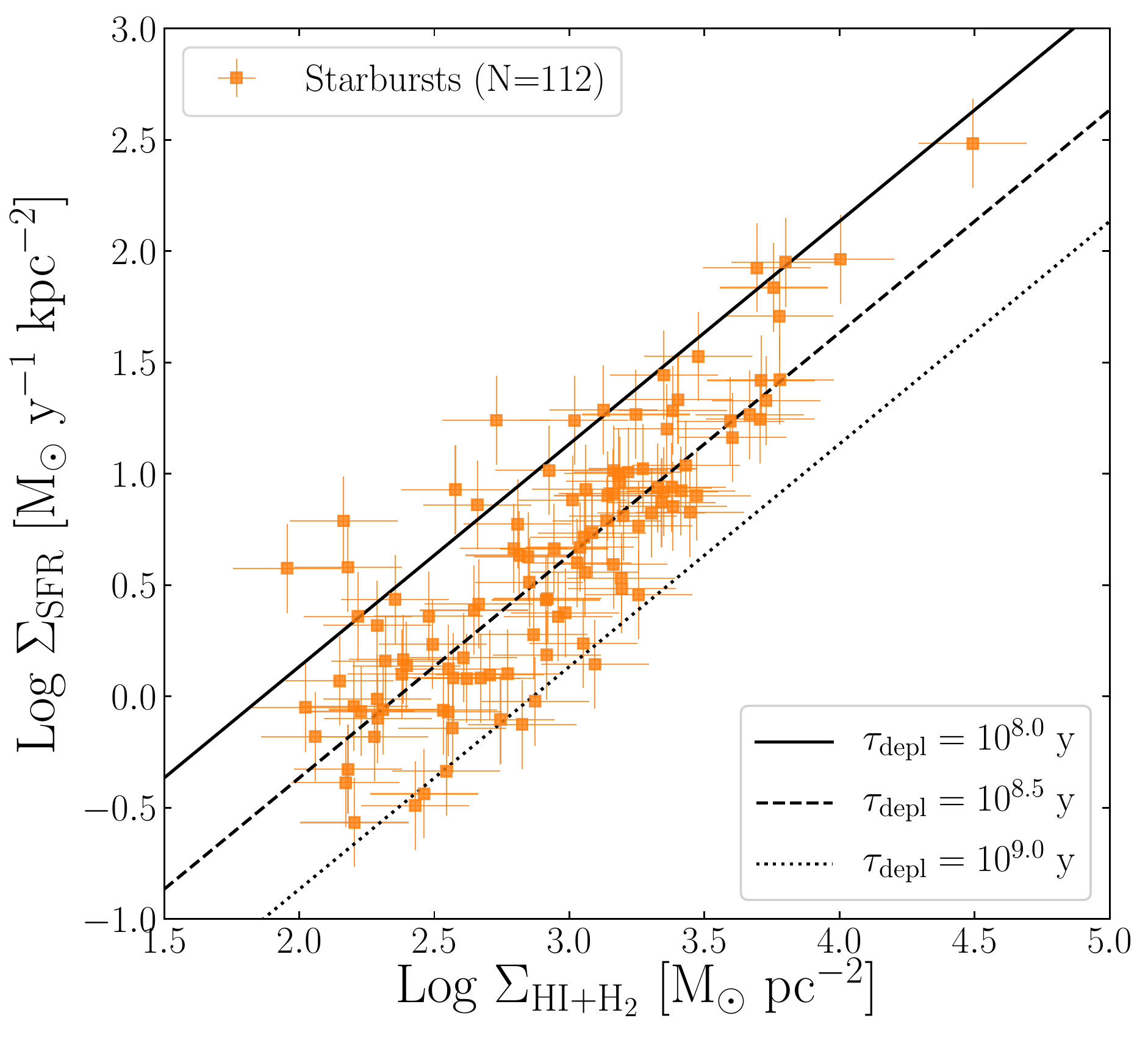}{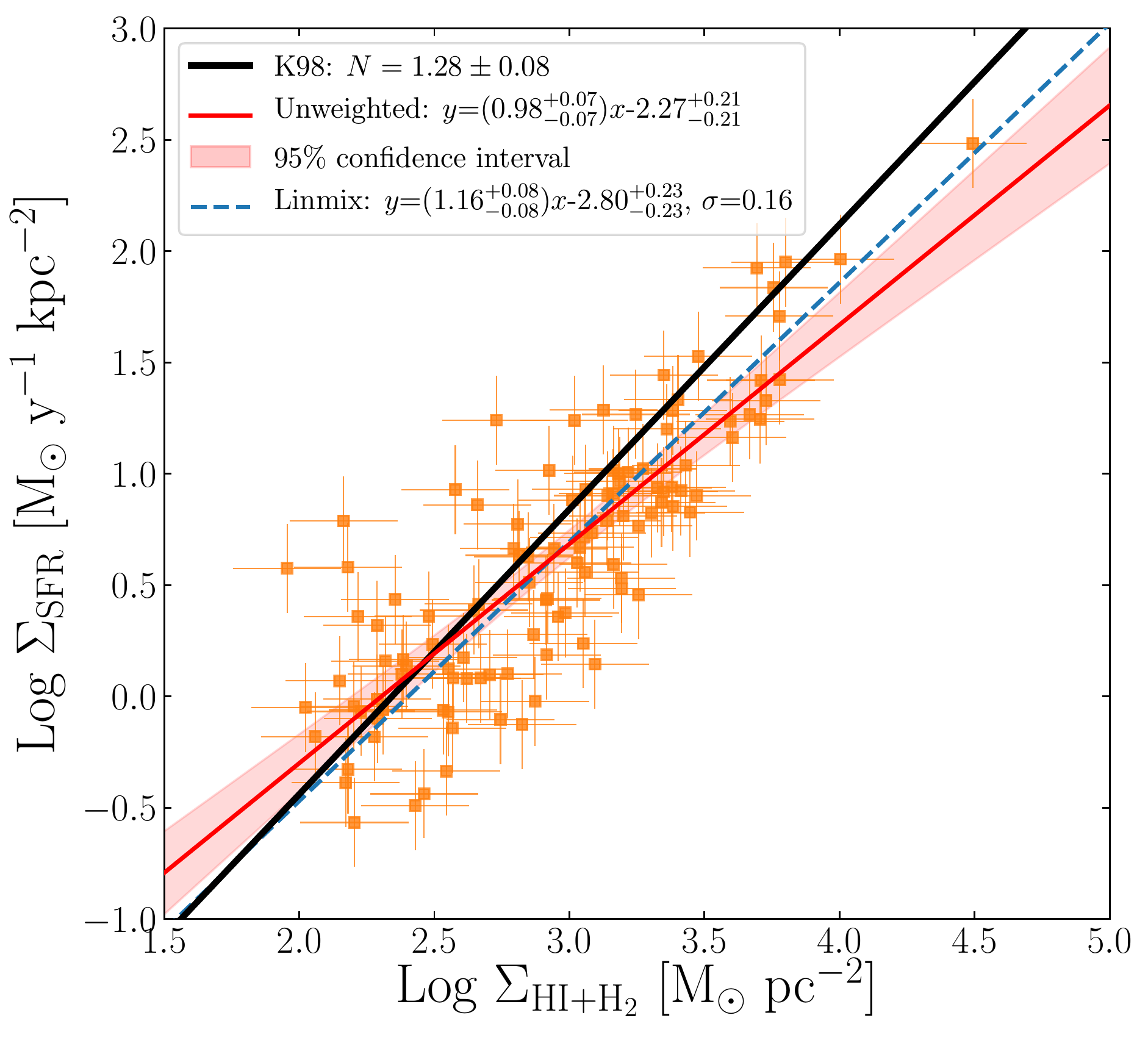}
    \caption{Schmidt law for the starburst galaxy sample.  The left panel overlays lines of constant molecular gas depletion times, and the right panel overlays maximum likelihood fits to the starburst galaxy sample, and the fit derived previously for a smaller sample by \citetalias{Kennicutt1998} (solid black line), adjusted to consistent choices for the IMF and CO-to-H$_2$ conversion factor.  All molecular surface densities were derived using a Milky Way value for \xco{}.  The starburst disks are all molecular dominated so the H$_2$ surface density is taken as proxy for the total gas surface density.}
    \label{fig:starburst_sample_only}
\end{figure*}

Figure~\ref{fig:starburst_sample_only} shows the dependence of the disk-averaged SFR surface density on mean gas (H$_2$) surface density for the starburst galaxy sample.  For this initial comparison a constant Milky Way value for \xco{} has been assumed.  The same data are plotted in both panels but the left panel shows lines of constant gas depletion time overplotted (including a correction factor of 1.36 for helium and heavy elements in the gas masses), while the right panel shows various types of linear fits to the current data. These fits, described in more depth in \citetalias{delosReyes2019}, are labeled ``unweighted'' and ``linmix.''
``Unweighted'' denotes ordinary least squares linear regression unweighted by errors, while ``linmix''\footnote{The linmix algorithm has been ported to a Python package by J. Meyers and is available on github at \url{https://github.com/jmeyers314/linmix}.} is a hierarchical Bayesian model developed by \citet{Kelly2007} that attempts to account for both $x$- and $y$-errors, as well as intrinsic random scatter.
We further discuss these methods in Appendix~\ref{appendix:stats}.
Along with the linear fits, the right panel of Figure~\ref{fig:starburst_sample_only} also displays the corresponding fit to starburst galaxies from \citetalias{Kennicutt1998}, but with the SFRs and molecular masses adjusted to incorporate the assumed IMFs and \xco{} conversions adopted for this paper.

Overall the relation shown in Figure~\ref{fig:starburst_sample_only} is similar to those derived previously by \citetalias{Kennicutt1998} and many subsequent authors \citep[e.g., ][]{Daddi2010, Genzel2010,Liu2015}.  The slope of the relation is approximately linear ($n =$ 0.98--1.16, depending on the fitting method employed).  This is somewhat shallower than found by \citetalias{Kennicutt1998} (1.28$\pm$0.08) but generally consistent with the shallower slopes found in the more recent studies.  Note that the choice of fitting algorithm has a significant effect on the derived slope; for consistency within this paper and with \citetalias{delosReyes2019} we generally will discuss the linmix hierarchical Bayesian fits as the defaults.  The median gas depletion time of 240 Myr is similar to that found by \citetalias{Kennicutt1998}, and is more than an order of magnitude lower than the average depletion time of 3.2 Gyr in the comparison sample of spiral galaxies from \citetalias{delosReyes2019}.  This is entirely consistent with the selection of the former as bursting systems by definition. The root-mean-square dispersion of galaxies about the fitted relation is approximately 0.33~dex, similar to \citetalias{Kennicutt1998}.  Although the currently derived slope differs from the fit in \citetalias{Kennicutt1998} by more than the statistical uncertainties, it is not entirely surprising in light of the small sample size and considerable uncertainties in the data used in \citetalias{Kennicutt1998}.

\subsection{Combined Star Formation Law for Starburst and Non-Starbursting Galaxies}

\begin{figure*}
    \centering
    \epsscale{1.1}
    \plottwo{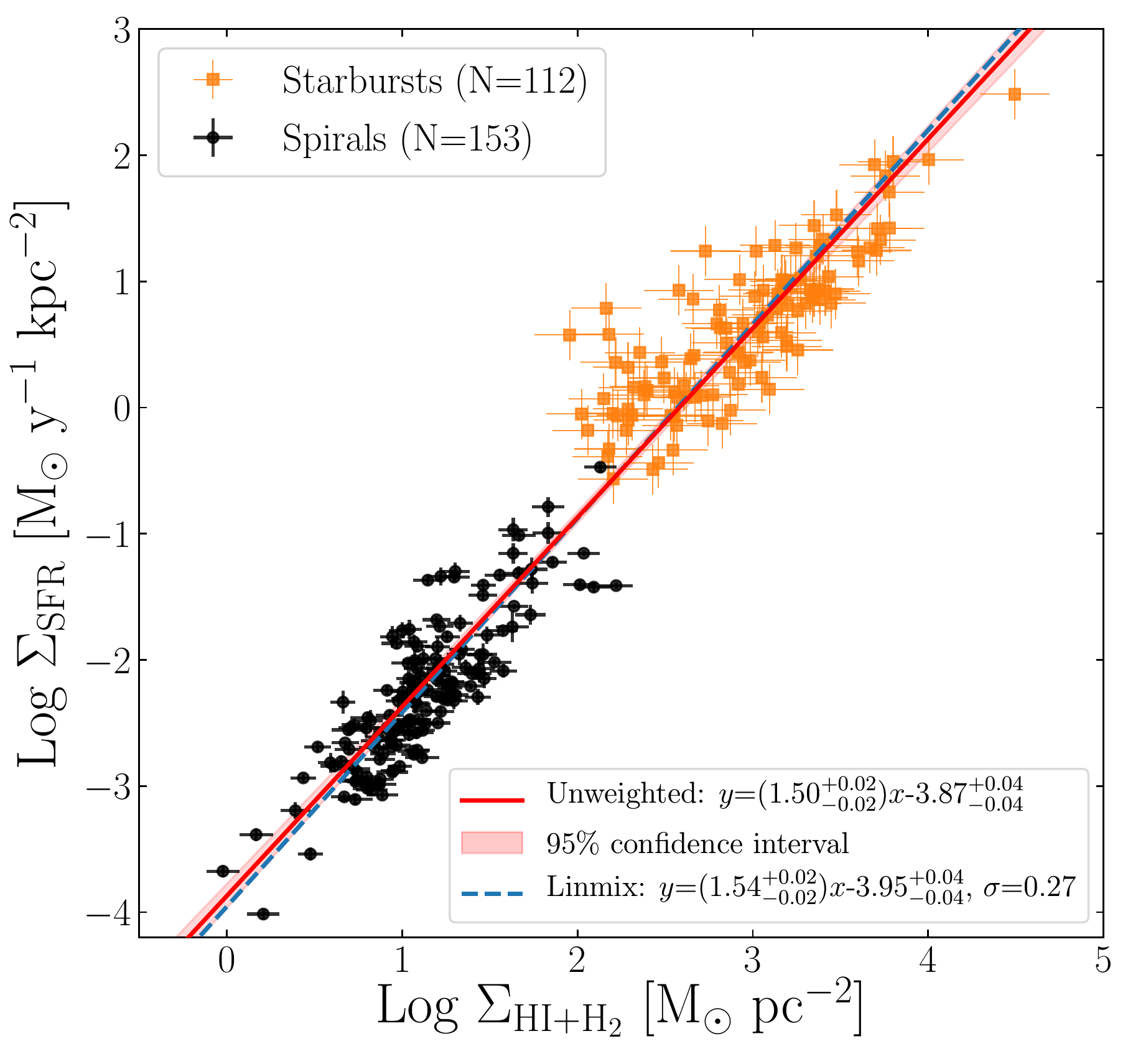}{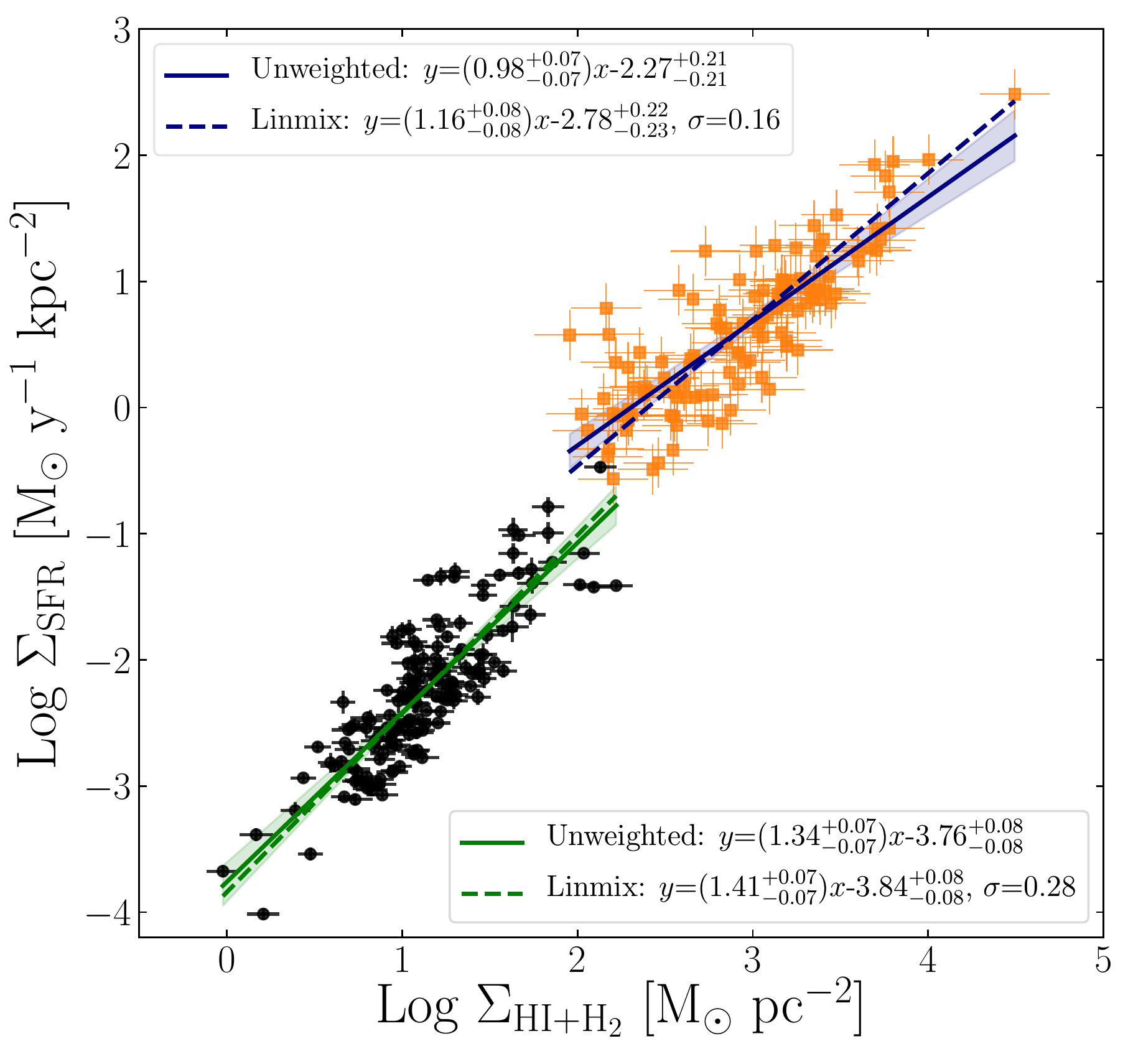}
    \caption{Relation between SFR surface density and total (atomic plus molecular) gas surface density for the the combined sample of normal spiral disks from Paper I (black points) and the starbursts in this sample (orange points). A constant Milky Way value for the CO-to-H$_2$ conversion factor is assumed for all galaxies.  Left:  A single power law is fitted to all of the data.  Right:  Separate power laws are fitted to the normal galaxies and the starburst galaxies. Shaded regions represent 95\% confidence intervals about the unweighted fits.}
    \label{fig:Schmidt_total_fits}
\end{figure*}

We next combine the data for the starbursts with the data from \citetalias{delosReyes2019} for normal spiral star-forming disks, and the results are summarized in Figure~\ref{fig:Schmidt_total_fits}.  We refrain from including the low surface brightness dwarf galaxies from \citetalias{delosReyes2019} at this stage, because they are drawn from a completely different galaxy population than any of the starburst galaxy hosts studied in this paper, and because data on molecular gas is absent for most of them.  In the left panel we fit the combined dataset to a common single power-law, as was done in \citetalias{Kennicutt1998}, whereas in the right panel we fit the non-starbursting spiral galaxies and the starbursts separately.

The left panel of Figure~\ref{fig:Schmidt_total_fits} shows that the observed surface densities are reasonably well fitted by a single Schmidt power law, now with a best fitted slope of $n=1.50\pm0.02$, as compared to $n=1.4\pm0.15$ for the much smaller sample in \citetalias{Kennicutt1998}.  We caution that the listed fitting uncertainties only include formal fitting errors; other observational and systematic errors across the samples (see \S\ref{sec:uncertainties}) may introduce larger uncertainties in the slope of the relation, and even the choice of fitting method affects the slope, as shown in Figure~\ref{fig:Schmidt_total_fits}.  Taking these additional terms into account the full uncertainty in the slope is estimated to be $\pm$0.05 dex.  The rms scatter of galaxies around the fitted power law is $\sim$0.37~dex.  
This is similar to the scatter about the relations fitted to normal disks and starbursts alone ($\sim$0.31 and 0.33~dex, respectively), and similar
to the dispersion in the combined relation of \citetalias{Kennicutt1998}.  The persistence of this factor-of-two dispersion despite the improved measurements suggests that the scatter in the Schmidt power law is physical in origin.  

The simple single power-law approximation is very widely applied in analytic studies and as a sub-grid prescription in numerical models and simulations.
The current results confirm that within the limitations of any such approximation, such a recipe remains a credible fit to the observed relation.  It is interesting that the overall slope $n \sim 1.5$ coincides with expectations for a simple picture in which the large-scale star formation rate is largely driven by self-gravitational time scales \citep{Elmegreen2015}. 

Closer examination of Figure~\ref{fig:Schmidt_total_fits}, however, reveals that when the normal spiral and starbursts are fitted separately they are not consistent with a common power-law fit.  In particular the starbursts appear to follow a significantly shallower relation with slope $n=1.98\pm0.07$ compared to $n=1.34\pm0.07$ for the non-starbursting spirals, as shown in the right panel.  The ability of a single power law to fit both sets of galaxies appears to be a fortuitous combination of the two different slopes, combined with a systematic shift in zeropoints between the relations for gas densities in region of overlapping surface densities ($\sim$100\,M$_\odot$\,pc$^{-2}$).  This shift is significant, $\sim$0.6 $-$ 0.7\,dex in SFR surface density at the transition gas surface density of $\sim$100\,M$_\odot$\,pc$^{-2}$, depending on the precise fitting method used.  Although the full range of star-forming galaxies is credibly fit by a common single power law, there is strong evidence in the data for a bimodal (or multimodal) relation, both in the slope and the zeropoint of the law.

\begin{figure}
    \centering
    \epsscale{1.1}
    \plotone{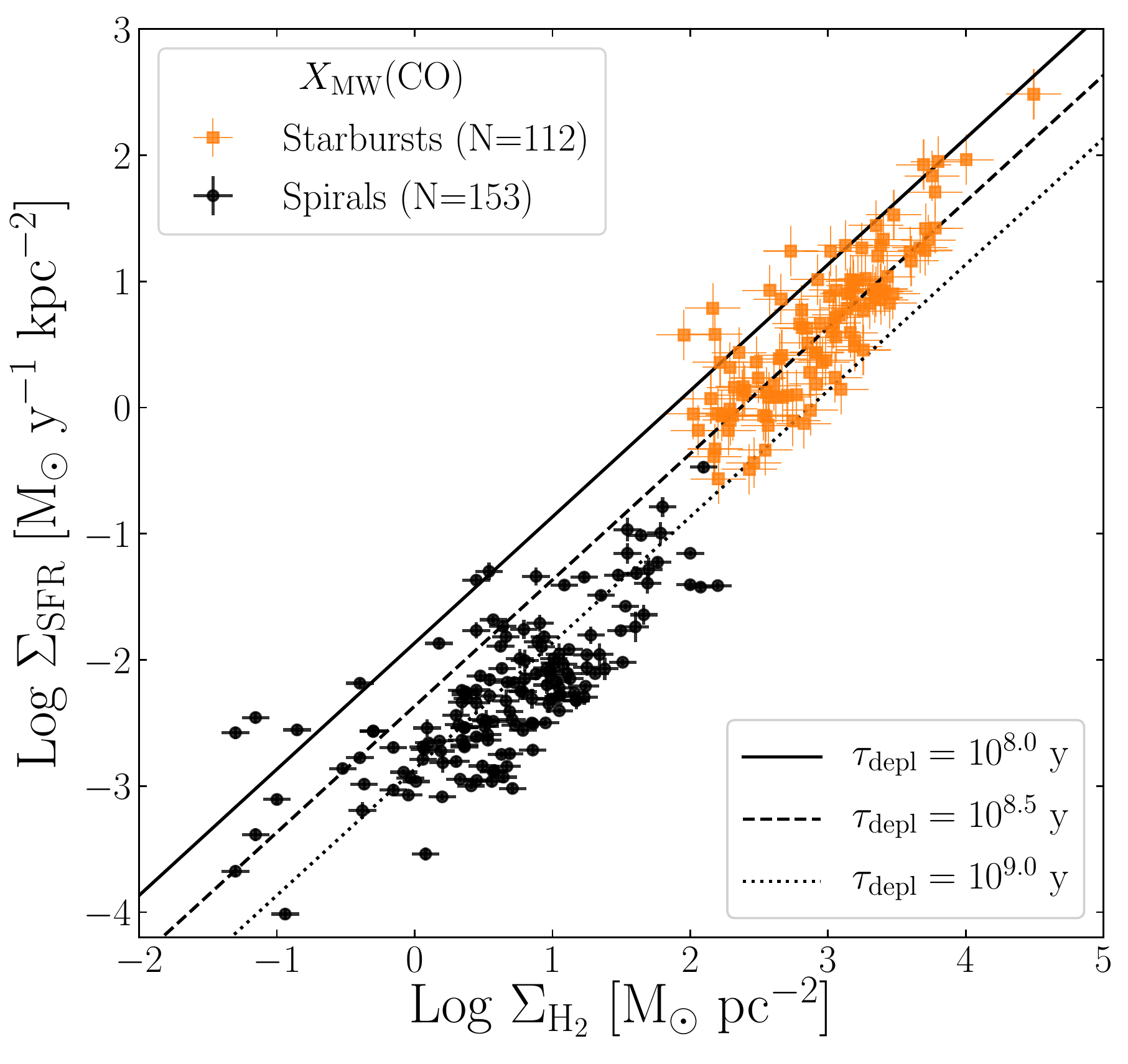}
    \caption{Relation between SFR surface density and molecular gas surface density for the the combined sample of normal spiral disks from Paper I (black points) and the starbursts in this sample (orange points).  A constant Milky Way value for the CO-to-H$_2$ conversion factor is assumed for all galaxies. Lines represent constant molecular gas depletion times.}
    \label{fig:Schmidt_H2}
\end{figure}

The pronounced change in the slope of the Schmidt law seen in Figure~\ref{fig:Schmidt_total_fits} largely disappears when the SFR surface density is plotted instead as a function of average molecular gas surface density, as illustrated in Figure~\ref{fig:Schmidt_H2}.  As already reported by \citet{delosReyes2019} and previously by numerous authors \citep[e.g.,][and references therein]{Leroy2013}, molecular surface density correlates roughly linearly with the SFR surface density even in normal spiral disks.  For the starbursts we are assuming that the molecular hydrogen dominates the gas surface density, but for the normal spiral sample molecular gas typically comprises only half or less of the gas density within the star-forming disks, so the net effect is to shift the latter points to the left (relative to Figure~\ref{fig:Schmidt_total_fits}) in the Schmidt relation.  For this comparison we adopt a constant Milky Way \xco{} conversion factor for all of the galaxies.  The slope of the molecular Schmidt law is close to linear even for the normal spiral galaxies \citep{delosReyes2019}. The offset is somewhat smaller than seen in the right panel of Figure~\ref{fig:Schmidt_total_fits}, $\sim$0.5 dex for the molecular law compared to 0.7 dex for the atomic$+$molecular relation, {\it when using the same CO-H$_2$ conversion factor}.  

\citetalias{Kennicutt1998} found that the global SFR surface densities of disks were also well correlated with the ratio of
gas surface density to dynamical time, taken in this case to be the orbital time scale of the star-forming disk (Silk-Elmegreen law):

\begin{equation}
    \Sigma_{\mathrm{SFR}} = A \frac{\Sigma_{\mathrm{gas}}}{\tau_{\mathrm{dyn}}}
    \label{eq:SE}
\end{equation}

\begin{figure}
    \centering
    \epsscale{1.1}
    \plotone{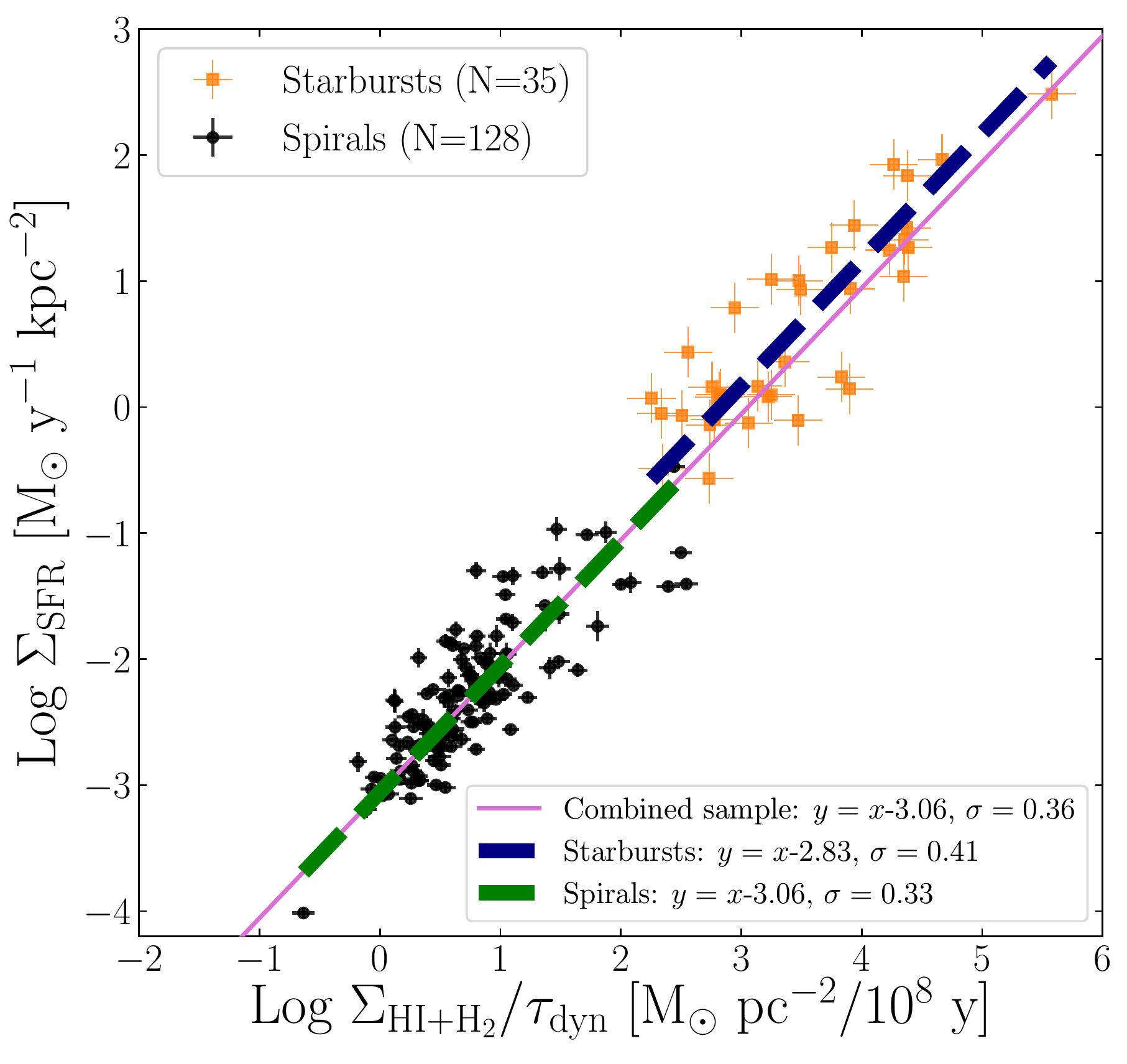}
    \caption{Correlation between SFR surface density and the ratio of total gas surface density to the 
    orbital period at the outer edge of the star-forming disks, or ``Silk-Elmegreen" law.  Normal and starbursts are indicated as in Figure~\ref{fig:Schmidt_total_fits}.  The solid points assume a fixed (Milky Way) value of the \xco{} conversion factor for all galaxies. 
    The pink solid line indicates the best-fit line with slope fixed to unity for the composite sample, while the blue and green dashed lines indicate the best-fit lines for the starburst and spiral samples, respectively. Root-mean-square dispersions are listed as $\sigma$.
    }
    \label{fig:Silk-Elmegreen}
\end{figure}

Figure~\ref{fig:Silk-Elmegreen} shows this relation for our composite sample of normal spiral and starburst galaxies.  In this case the number of starburst galaxies is much smaller, because reliable disk rotation times could only be measured for galaxies with well-resolved CO disks and evidence for rotation-dominated motions.  The data points all assume a constant Milky Way \xco{} conversion factor for both the normal spiral and starburst galaxies, and the superimposed line is a median fit to those data, with a fixed slope of unity as dictated by Equation~\ref{eq:SE}.  As was reported earlier in \citetalias{Kennicutt1998}, this Silk-Elmegreen correlation is competitive with the Schmidt law as an empirical descriptor of the star formation law.  Some evidence for a zeropoint shift between spirals and starbursts is evident here as well, however, as shown by the dashed lines, with a mean offset of 0.2 dex.

\subsection{Dependence on Assumed CO/H$_2$ Conversion Factors}

\begin{figure*}
    \centering
    \epsscale{1.1}
    \plottwo{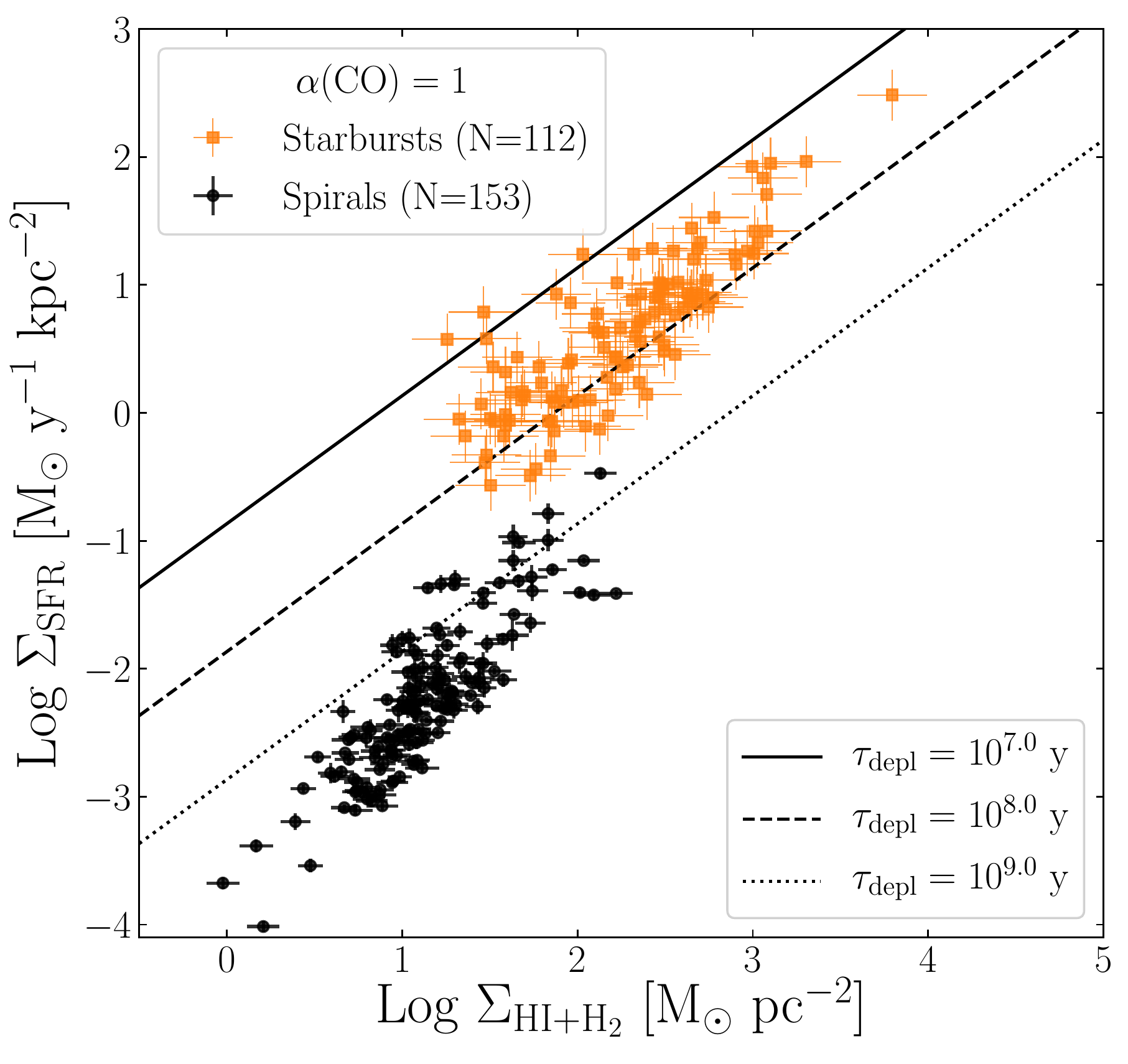}{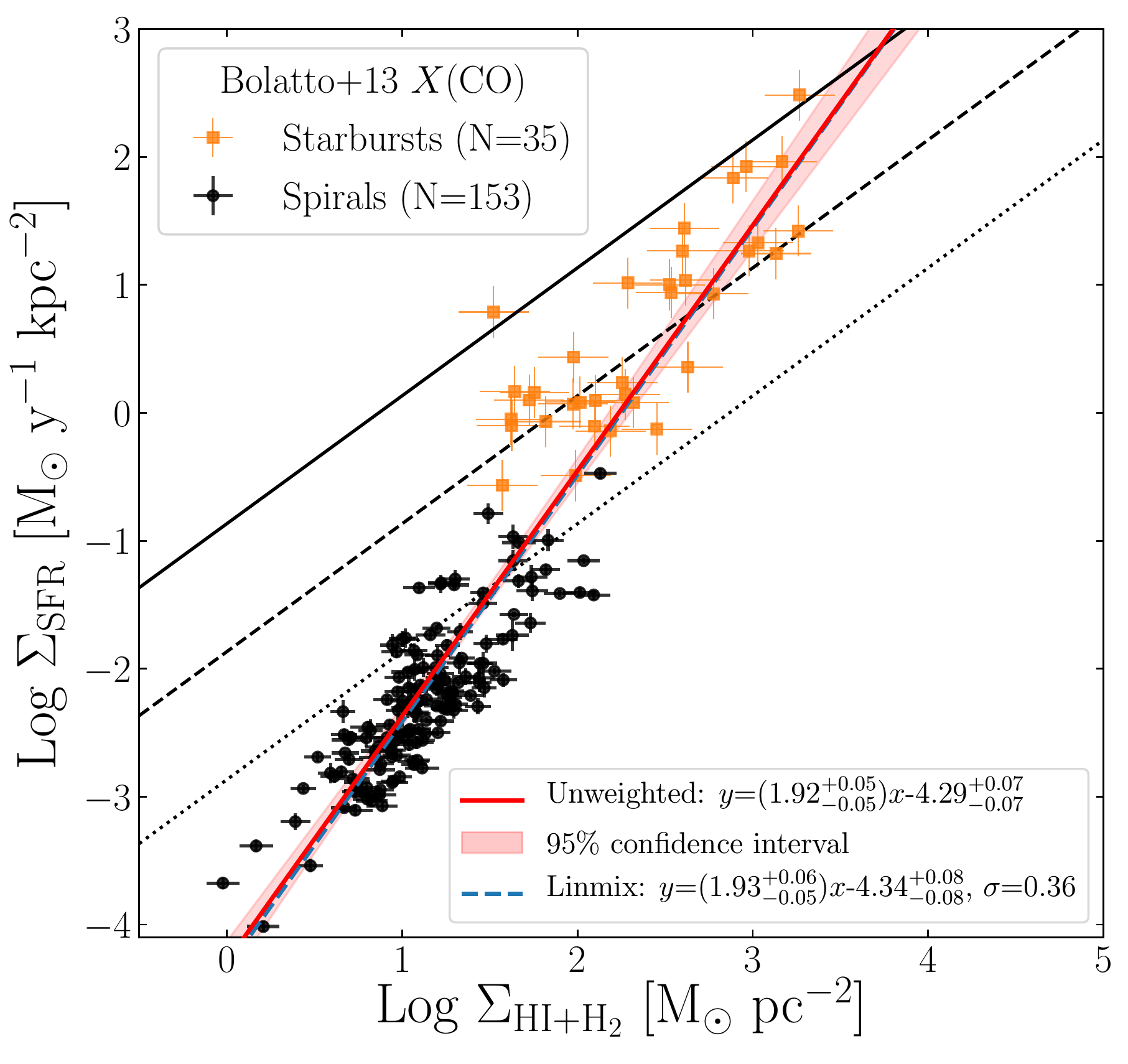}
    \caption{Left:  The combined star formation law (as a function of total gas surface density) when a CO conversion factor \xco\ of 0.2 times the Milky Way value is applied to the starbursts (while maintaining the Galactic conversion factor for the normal spirals).  (The spiral sample has the same fit as in Figure~\ref{fig:Schmidt_total_fits}.)  Right:  The composite Schmidt law with molecular gas masses and surface densities calculated using the X(CO) prescription in Eq.~(5) taken from \citet{Bolatto2013}. This prescription produces progressively lower molecular masses per unit CO intensity as a function of surface density, and a roughly quadratic Schmidt law ($n\sim2$). Thin black lines mark lines of constant depletion time, of (top to bottom) 10 Myr, 100 Myr, and 1 Gyr, respectively.}
    \label{fig:Lower_xco}
\end{figure*}

Although we have arbitrarily applied a single value for the \xco{} conversion factor throughout our analysis up to now, considerable evidence in the literature suggests that the molecular gas in luminous starbursts is considerably warmer than in the clouds which dominate the emission in normal spiral disks, which will tend to increase the emissivity of the CO lines for a given H$_2$ column density \citep[][and references therein]{Bolatto2013}.  Gas in the starburst regions may also have higher velocity dispersions, which would also tend to lower \xco{}.  Different authors have applied a wide range of values, but most argue for a conversion factor that is 3--7 times lower than for the Milky Way.  In the left panel of Figure~\ref{fig:Lower_xco} we follow the recommendations in \citet{Bolatto2013}, and apply a value \xco{} = 0.2 $X_{\mathrm{MW}}$ ($\alpha_{\mathrm{CO}}$ = 0.86) to the starbursts, while maintaining the Milky Way conversion factor for the non-starbursting spiral galaxies.  The net effect (trivially) is to shift the starbursts to left by 0.7 dex relative to their positions in Figure~\ref{fig:Schmidt_total_fits}, thereby producing a very strongly bimodal Schmidt law, consistent with the results reported previously by \citet{Daddi2010} and \citet{Genzel2010}, when considering the assumptions made about \xco{}.  Note however that most of these shift result from the application of bimodal values for \xco{}. 

It seems unlikely that only two discrete values of \xco{} apply to the galaxies in our study.  As reviewed by \citet{Bolatto2013}, it is more likely that \xco{} varies continuously across the wide range of physical environments found in these galaxies.  
As a test of the sensitivity of our results to such scalings of \xco{} we adopted an algorithm recommended (with many qualifications) by \citet{Bolatto2013}:

\begin{equation}
    \alpha_{\mathrm{CO}} = 2.9 \exp\left(\frac{0.4}{Z^{'} \Sigma^{100}_{\mathrm{GMC}} }\right) \left(\frac{\Sigma_{\mathrm{total}}} {100~M_{\odot}~\mathrm{pc}^{-2}}\right)^{-\gamma} 
    \label{eq:bolatto}
\end{equation}
where $\gamma = 0$ for total (stellar+gas) surface densities $\Sigma_{\mathrm{total}} < 100~M_\odot~\mathrm{pc}^{-2}$ and 0.5 for higher $\Sigma_{\mathrm{total}}$. The conversion factor $\alpha_{\mathrm{CO}}$ scales directly with \xco{} and for the assumptions applied in this paper has the value 4.0 for a Milky Way factor. 
The first exponential term on the right side of Equation~\ref{eq:bolatto} allows for a dependence on metal abundance ($Z'$), which we ignore by assuming approximately solar composition for all galaxies, and on the typical surface density of molecular clouds in units of $100~M_{\odot}~\mathrm{pc}^{-2}$ ($\Sigma^{100}_{\mathrm{GMC}}$), which we assume to be unity.

Since the molecular gas density on the right side of the equation depends on $\alpha_{\mathrm{CO}}$ on the left side, the equation usually needs to be solved iteratively.  However, since it is nearly impossible to derive reliable stellar surface densities for the LIRGs and ULIRGs anyway, we opted instead to use the measured dynamical masses (for regions where this could be measured reliably) to derive total surface densities. We also assumed solar metallicity by default for all galaxies.  The resulting star formation relation is shown in the right panel of Figure~\ref{fig:Lower_xco}.  The net effect of Equation~\ref{eq:bolatto} is to apply a decreasing \xco{} factor with increasing surface density, and thereby steepen the slope of the resulting Schmidt law.  Applying this prescription to our data yields a nearly quadratic relation {\bf ($n = 1.92 \pm 0.05$)}.  Our results confirm an earlier study by \citet{Narayanan2012}, who found a similarly steep relation albeit with a somewhat different but physically-motivated prescription for \xco{}.  

One other final way to consider the apparent bimodality is to ask what value of \xco{} would need to be applied to the starburst galaxies to cause their Schmidt relation to merge smoothly with that for the normal spirals at the transition surface density of $\sim$100\,M$_{\odot}$\,pc$^{-2}$?  Since the slope of the starburst law is approximately linear and the offset in $\Sigma_{{\rm{SFR}}}$ is $\sim$0.6\,dex, \xco{} for the bursts would need to be $\sim$4 times {\it higher} than for the Milky Way.  This trend is opposite to the usual assumptions about \xco{} in starbursts, and the only conceivable way to produce such a high value would be if the LIRGs and ULIRGs were much more metal poor than normal spirals, by factors of 4--10 depending for one's assumptions about the scaling of \xco{} with metallicity \citep{Bolatto2013}.  Spectroscopic studies by \citet{Rupke2008} and \citet{Rich2012} suggest indeed that the central regions of LIRGs and ULIRGs may be underabundant in metals by up to a factor of two, so this factor could account for some but certainly not all of the offset seen here.

\subsection{Tests for Spurious Sources of Bimodality}

Before we discuss these results further it is important to rule out any possible spurious sources of bimodality, resulting from the way in which the galaxy samples were defined or measured.  Here we summarize tests which were applied to rule out such origins.

\subsubsection{Direct CO Mapping {\it vs} Indirect Measurements}

\begin{figure*}
    \centering
    \epsscale{1.1}
    \plottwo{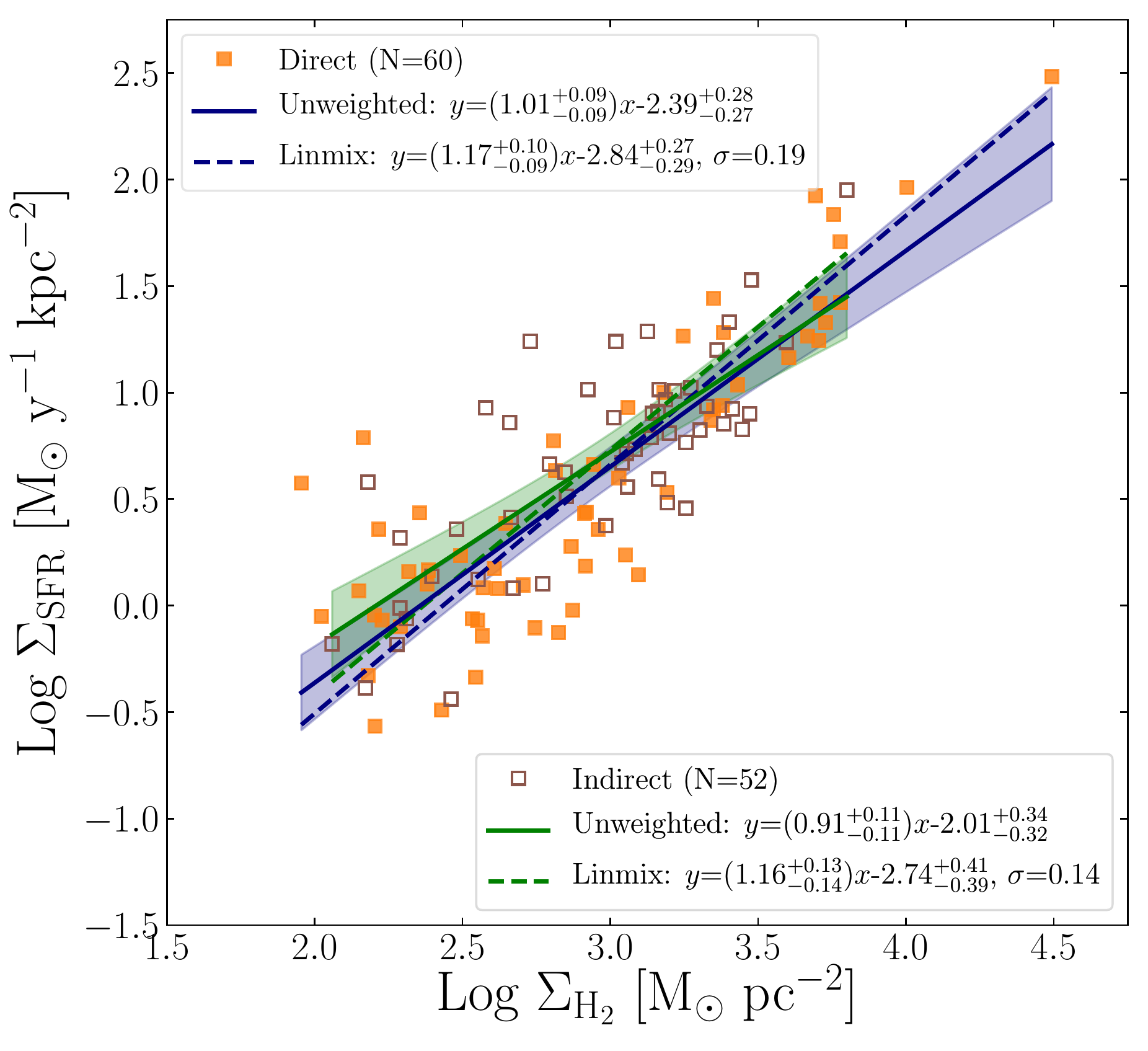}{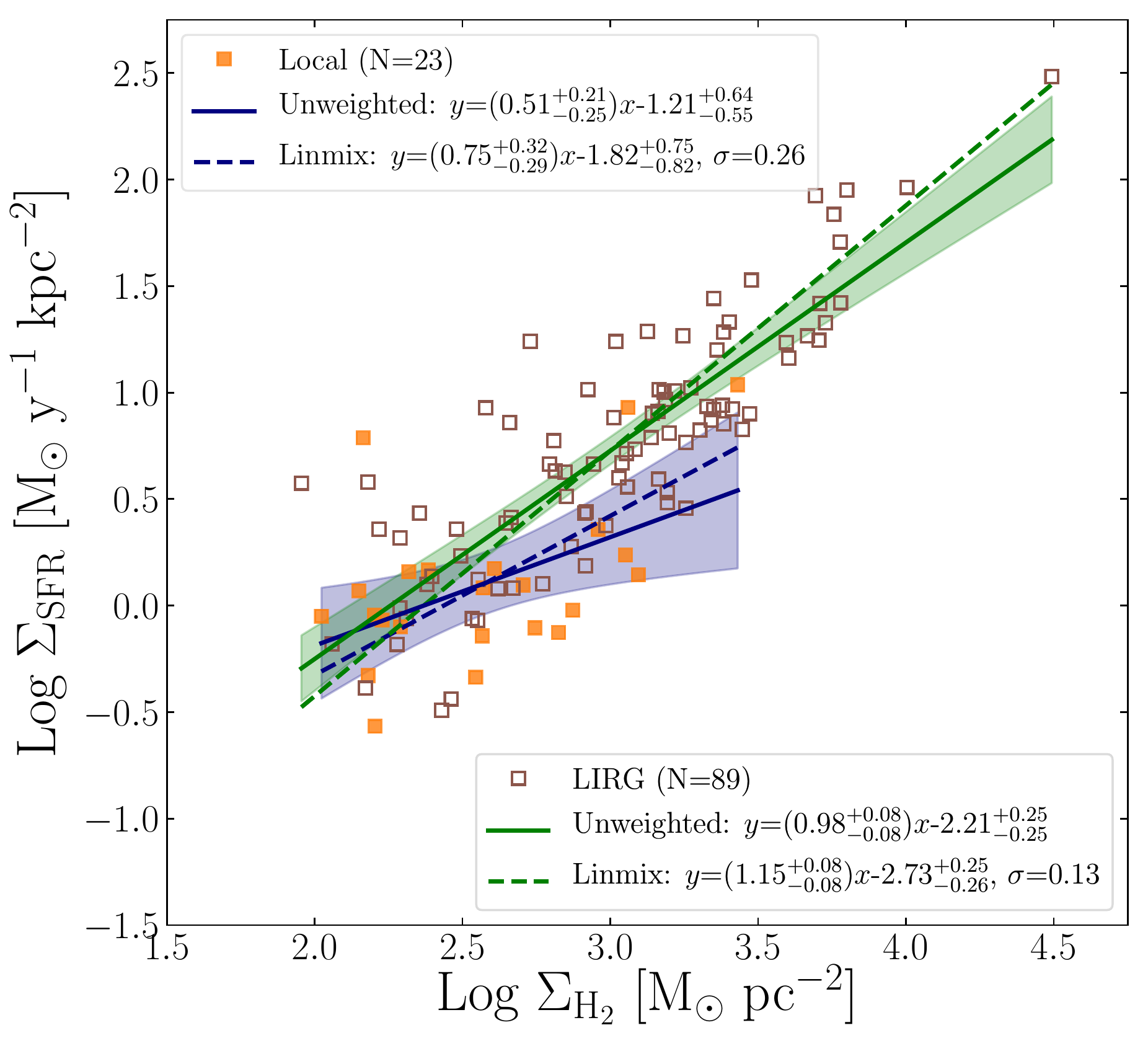}
    \caption{The Schmidt law for starburst galaxies in our sample, subdivided in the left panel between disks with direct measurements of molecular disk sizes (solid points) and indirect measurements (open points), and in the right panel between local, well-resolved circumnuclear disks (solid points) and more distant LIRGs and ULIRGs (open points). Shaded regions represent 95\% confidence intervals about the unweighted fits.} 
    \label{fig:starburst_subsamples}
\end{figure*}

As we have discussed in ~\ref{sec:diameters} and will demonstrate further in Section ~\ref{sec:literature}, the adoption of physically unrealistic sizes for the star-forming disks can introduce systematic biases in the resulting Schmidt laws.  In this study, approximately half of the starbursts have molecular disk sizes measured directly in CO from interferometer maps, while for the others the diameter is inferred directly from infrared or emission-line disk sizes.  In order to test whether these could be influencing the form of the star formation law, we show in the left panel of Figure~\ref{fig:starburst_subsamples} the Schmidt relations for starbursts with directly measured CO disks (solid points) along with those with indirect size measurements (open points).  The best-fit relations for the two subsets of galaxies agree within the fitting uncertainties, both in slope and zeropoint.  This appears to rule out systematic errors in our diameter measurements as being a significant source for the observed shift in star formation laws.  The result also offers reassurances that issues with missing flux in the interferometer data, already checked during the data compilation process, are not a significant issue for these data.

\subsubsection{Luminous and Ultraluminous Starbursts {\it vs} Bar-Driven Circumnuclear Disks}

As discussed in \S\ref{sec:data} our starburst galaxies were drawn from two parent samples: luminous and ultraluminous infrared starburst galaxies selected on the basis of their high SFRs (L$_{\mathrm{IR}} \ge 10^{11}\,L_\odot$ or SFR $\ge 16\,M_\odot\,\mathrm{yr}^{-1}$), and lower-luminosity circumnuclear disks, mainly located in strongly barred galaxies, or less luminous instances of the accretion-driven starbursts (e.g., M82).  Apart from the differences in selection and SFR, the LIRGs and ULIRGs tend to be more distant and less well resolved than the local circunuclear starbursts. 

The right panel of Figure~\ref{fig:starburst_subsamples} compares the SFR and gas surface densities for these two subsamples.
While the ``LIRG'' sample (empty points, green lines) appears nearly identical to the relation found for the entire sample (Figure~\ref{fig:Schmidt_total_fits}), the ``local'' circumnuclear disks (filled points, blue lines) appear to follow a shallower power law, and with the average SFR per unit gas mass shifted to somewhat lower values ($\sim$0.2 dex) on average.  In view of the small number of local starbursts (23), the narrower range in surface densities, and the large uncertainties in the fit for these galaxies it is difficult to ascertain whether these differences are physically significant.  Later in the discussion section we consider whether some of the difference could be real and point to an intermediate regime between normal spiral galaxies and the luminous and ultraluminous starbursts.  In either case, however, both sets of starbursts are strongly offset from the Schmidt law seen in non-starbursting disks.

\subsubsection{Calibrations of SFRs and Gas Densities}

Different prescriptions were used to measure the SFRs of the normal and starburst galaxies (reflecting the different stellar population ages and dust attenuation regimes in the two subsamples). Could this introduce a spurious break in the star formation law?  

For the starbursts the SFRs were estimated directly from their infrared luminosities, on the assumption that virtually all of the young starlight is reprocessed by dust.  The SFR calibration assumes furthermore an average age for the dust-heating population of 50 Myr.  On the other hand, the lower typical dust opacities of the extended normal spiral disks ($<$1 mag on average) mean that their SFRs needed to be measured from a combination of UV and IR fluxes in order to correct for dust attenuation (see \citetalias{delosReyes2019}, which used the combination of 155\,nm ultraviolet fluxes from GALEX and 22$-$25\,$\mu$m infrared fluxes from IRAS, Spitzer, or the WISE surveys).  The spiral SFR calibrations also assume steady-state star formation over the past Gyr, to properly incorporate additional dust heating from evolved stars in these galaxies.  Clearly there is a range of star formation histories and dust geometries across both the normal and starburst samples, so neither of these SFR prescriptions is expected to be precise for individual objects; indeed these variations are likely to be a significant source of the observed dispersion in the Schmidt law.  

We tested the sensitivity of the assumptions by applying both SFR calibrations to the same starburst regions, and we find that they generally give consistent SFRs to within $\sim$30\% with no systematic offset.  The reason appears to be that the starbursts generally exhibit strongly elevated mid-infrared emission in the 24\,$\mu$m region which largely compensates for the relative under-weighting of the infrared luminosity in the calibration for normal star-forming disks.  In any case, any systematic shifts in SFRs due to these population and calibration differences are much too small to account for the $\sim$0.7\,dex offset in the star formation relations.  Likewise, changing the assumed ages of the starbursts within plausible limits only lowers the resulting SFRs by $\ll$30\% \citep{Leitherer99}, again much less than the five-fold offset seen in the star formation relations.

Based on these various tests, we tentatively conclude that the origin for the break in star formation laws cannot be ascribed to any systematic differences in methods used to measure the gas masses, SFRs, or sizes of the star-forming regions.

\section{A Consistency Test Using Dust Surface Densities}
\label{sec:dustSchmidt}

We have seen that the interpretation of the form of the star formation law is critically dependent on the assumed behavior of the CO-to-H$_2$ conversion factor in different star formation environments, confirming earlier results from \citet{Daddi2010} and \citet{Genzel2010}.  Uncertainties in the behavior of \xco\ affect other studies of the molecular gas contents and depletion times in galaxies, and this has led researchers to explore alternative approaches to measuring or at least constraining the gas contents of galaxies.  One promising approach is to use interstellar dust masses as proxies for molecular gas masses \citep{Tacconi2020}. Thanks to the {\it Spitzer Space Telescope} and {\it Herschel Space Observatory} missions, well-measured mid-infrared to submillimeter spectral energy distributions (SEDs) are available for large samples of nearby normal spiral galaxies and infrared-luminous starburst galaxies, making it possible to measure far more robust dust masses than was previously possible.  Converting dust masses to estimated total gas masses requires the assumption of a dust/gas ratio (either fixed or with some parametric variation), but so long as one avoids metal-poor (and dust-poor) galaxies the systematic uncertainties due to variations in dust fractions are generally much smaller than those plaguing CO-based mass measurements.  

As a consistency check on the results in \S\ref{sec:results} we have analyzed the dependence of the SFR surface densities of a subset of our galaxies on the mean surface densities of {\it dust} in the same regions.  Since we are only attempting to examine the continuity of the Schmidt law between normal and starburst galaxies there is no need to convert from dust to gas masses; we only aim to compare the structural behavior of the SFR vs dust density relation to those found for gas in Figures~\ref{fig:Schmidt_total_fits} and~\ref{fig:Lower_xco}.  

We restricted our experiment to galaxies with published dust masses in the literature.  As discussed in \S\ref{sec:dust}, dust masses for 76 starburst galaxies in the GOALS sample were taken from \citet{Shangguan2019}.  These are based on infrared SEDs fitted using the models of \citet{DraineLi2007} (\citetalias{DraineLi2007}), including any necessary aperture corrections as described in \S\ref{sec:IR} and \S\ref{sec:dust}.  Our comparison sample of non-starbursting spirals consists of 75 galaxies in the \citetalias{delosReyes2019} sample with dust masses measured from the KINGFISH project \citep{Aniano2020} and/or the {\it Dustpedia} project \citep{Clark2018}, a large compilation of multi-wavelength measurements of nearby galaxies observed with {\it Herschel}.  

Care was taken to place the three sets of dust masses on a consistent normalization.  All were measured using \citetalias{DraineLi2007} dust models, but the KINGFISH masses are based on an updated zeropoint calibration of the dust masses described by \citet{Draine2014}.  The re-scaled dust masses are 30\%\ lower than those based on the original \citetalias{DraineLi2007} calibration \citep{Aniano2020}, which we confirmed by comparing the \citet{Aniano2020} KINGFISH masses with those published previously by \citet{Draine2007} using the original calibration (average scaling 0.72 $\pm$ 0.03) and derived independently using the same calibration by the Dustpedia group (0.67 $\pm$ 0.02).  Dust surface densities were measured by normalizing the masses to the same surface areas discussed in \S\ref{sec:diameters}.
The dust surface densities for the starburst galaxies are listed in Table~\ref{tab:derived}, while Table~\ref{tab:dust} lists the dust surface densities for the non-starbursting spirals.

\begin{deluxetable}
{llc}
\tablecolumns{3} 
\tablecaption{Dust surface densities for spiral galaxies from \citetalias{delosReyes2019}. \label{tab:dust}} 
\tablehead{ 
\colhead{N} & \colhead{ID} & \colhead{$\log \Sigma_{\mathrm{dust}}$} \\
\colhead{} & \colhead{} &
\colhead{($[M_{\odot}~\mathrm{pc}^{-2}]$)}
}
\startdata
1 & MESSIER 074 & -0.99 \\
2 & NGC 0925 & -1.41 \\
3 & NGC 1291 & -1.59 \\
4 & NGC 1365 & -1.21 \\
5 & IC 0342 & -0.81 \\
6 & NGC 1566 & -1.29 \\
7 & NGC 1800 & -0.31 \\
8 & NGC 2146 & -0.54 \\
9 & NGC 2273 & -1.33 \\
10 & NGC 2403 & -1.08 \\
\enddata
\tablecomments{Only a portion of Table~\ref{tab:dust} is shown here; it is published in its entirety in the machine-readable format online.}
\end{deluxetable}

As expected the resulting dust masses and surface densities are strongly correlated with the corresponding gas masses and densities, with rms dispersions in $\log{G/D}$ of 0.19 and 0.26 dex for the normal spiral and starburst samples, respectively.  The median gas/dust ratios (including helium and heavy elements for gas) for the two samples are the same within these uncertainties: 138 for the local spirals and 129 for the starburst galaxies.  This assumes a standard Milky Way \xco{} conversion factor for all galaxies.  

\begin{figure}
    \centering
    \epsscale{1.1}
    \plotone{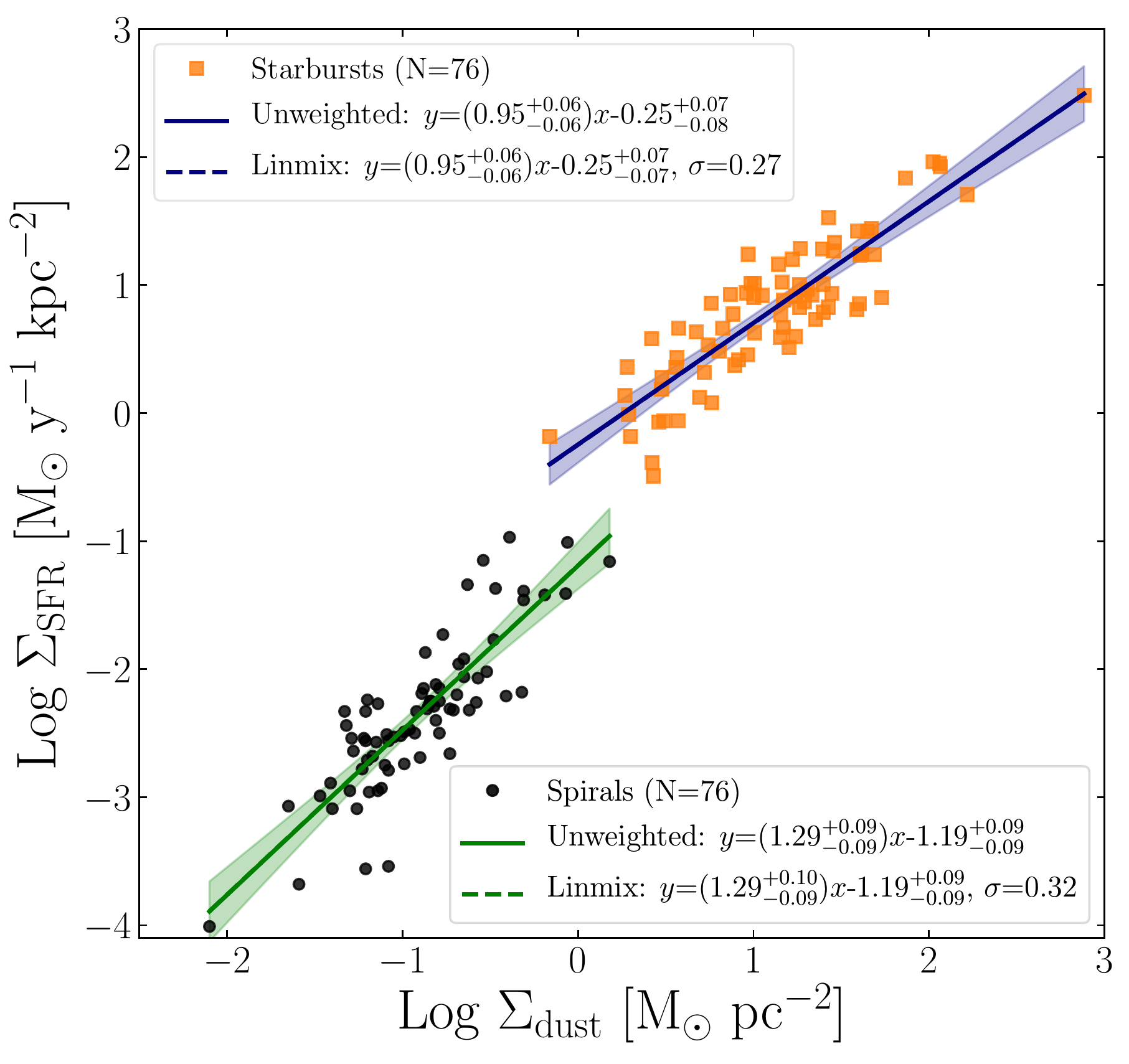}
    \caption{Correlation between SFR surface density and the surface density of interstellar dust for spiral (black points) and starbursts (orange points).  The results closely replicate those shown in the right panel of Figure~\ref{fig:Schmidt_total_fits} for SFR vs total gas surface densities, when a Milky Way value for the \xco{} conversion factor is applied. Shaded regions represent 95\% confidence intervals about the unweighted fits.}
    \label{fig:Schmidt_dust}
\end{figure}

Figure~\ref{fig:Schmidt_dust} shows the resulting composite relation between SFR surface density and \emph{dust} surface density, with separate fits to the spirals and starbursts.  The behavior of the plot is very similar to that seen in the SFR vs gas density relation shown in the right panel of Figure~\ref{fig:Schmidt_total_fits}, with a turnover in slope between normal spirals and circunuclear starbursts ($n =$ 1.29 to 0.95, respectively), and an offset to higher star formation efficiencies in the starbursts of $\sim$0.6 dex. This result reinforces the evidence presented earlier for a break in the Schmidt law between the two sets of galaxies, and demonstrates that the break cannot be a mere artifact of \xco{} conversion issues, because the dust surface densities are completely independent from the CO and (for normal spirals) \hi{} gas measurements.  Another interesting result is that the dust relation is most consistent with the earlier SFR vs gas surface density relation when a consistent Milky Way CO conversion factor is applied to all of the objects. 

We reiterate that the analysis above is only a pilot study, and readers should be aware of a few caveats before overinterpreting Figure~\ref{fig:Schmidt_dust}.  The derived dust masses fundamentally are dependent on the dust luminosity and temperature, but given that both the dust masses and SFRs are derived from the same infrared spectral energy distributions some co-variance between the two parameters in Figure~\ref{fig:Schmidt_dust} will be present, and the dispersion in the relation is probably supressed relative to the CO-based Schmidt laws.  This co-dependency is insufficient to account for the strong bimodality seen in the relation.  We intend to explore these issues further in 
Paper III.

Apart from offering insights into the form of star formation law, the results above and in \S3 offer some additional constraints on the behavior of the \xco{}\ conversion factor in the luminous starbursts.  As reviewed recently by \citet{Tacconi2020}, recent large-scale surveys of CO and dust emission in nearby and distant galaxies have cast some doubt on the application of grossly lower-than-Galactic values of xco{}\ values to the starbursts.  The results in this paper provide somewhat broad but hard bounds on the minimum and maximum bounds on \xco{}\ in these systems. Values significantly above \xco{$_{MW}$}\ are ruled out because the CO-inferred masses of the regions often equal or exceed the dynamically-inferred masses, confirming many previous studies \citep[e.g.,][and references therein]{Bolatto2013}.
At the same time, crude {\it lower} limits on \xco\ can be derived by comparing the resulting molecular gas depletion times with the dynamical timescales of the merging galaxies themselves.  As illustrated in Figure~\ref{fig:Lower_xco}, the adoption of very low values of \xco{} in the most luminous starbursts often produces uncomfortably short gas depletion times of as little as 10 Myr.  These depletion times are an order of magnitude or more shorter than the dynamical timescales of the host mergers, presumably during which the circumnuclear gas disks were assembled. The combination of these dynamical arguments would appear to confine the range of plausible values of \xco{} in the starbursts to within factors of less than a few of the Milky Way value.  As mentioned earlier the comparison of dust and gas masses in the galaxies is also consistent with realistic dust-to-gas ratios if \xco{} is not dramatically different from the Galactic value.

\section{Comparison to Previous Work}
\label{sec:literature}

Some of the results summarized above have already been reported or suggested in previous studies.  Systematically shorter gas depletion times and molecular depletion times in particular were observed in some of the first surveys of cold gas in normal and starburst galaxies \citep[e.g.,][]{Solomon1988} and in \citetalias{Kennicutt1998}, and perhaps most convincingly in the study of molecular depletion times of a diverse galaxy sample by \citet{Saintonge2011}. Those authors reported a continuous range in depletion times as a function of SFR per unit mass (specific SFR), with normal spiral galaxies and starbursts of the types studied here falling in the long and short extremes, respectively, of this range of the depletion times.

The apparent bimodality between the star formation laws for normal disks and starbursts, especially apparent when applying a bimodal \xco{} conversion factor, has also been reported and analyzed in previous papers, most notably by \citet{Daddi2010} and \citet{Genzel2010}.  Our results if anything show an even stronger bimodality, with a separation seen even when a constant \xco{} conversion factor is applied uniformly across both normal and starburst samples.  

Since the publication of \citetalias{Kennicutt1998}, a non-linear composite global star formation law has been reproduced in other studies, most recently by \citet{Liu2015}.  The latter report a significantly shallower composite slope ($n\sim1.2$) than we find in this study ($n\sim1.5$).  We have compared the two sets of results and believe that the main origin of the difference is in the use of radio continuum derived diameters for the star-forming disks in \citet{Liu2015}.   We compared the radio diameters with measured diameters of the CO and infrared disks for galaxies in common between our samples, and found that the radio diameters tend to be systematically smaller in the starburst galaxies, by nearly an order of magnitude on average (with a large scatter).  The origin for this offset is not completely clear, but may include contamination in the radio by AGNs \citep[as suggested already by][]{Liu2015}.

\begin{figure}
    \centering
    \epsscale{1.15}
    \plotone{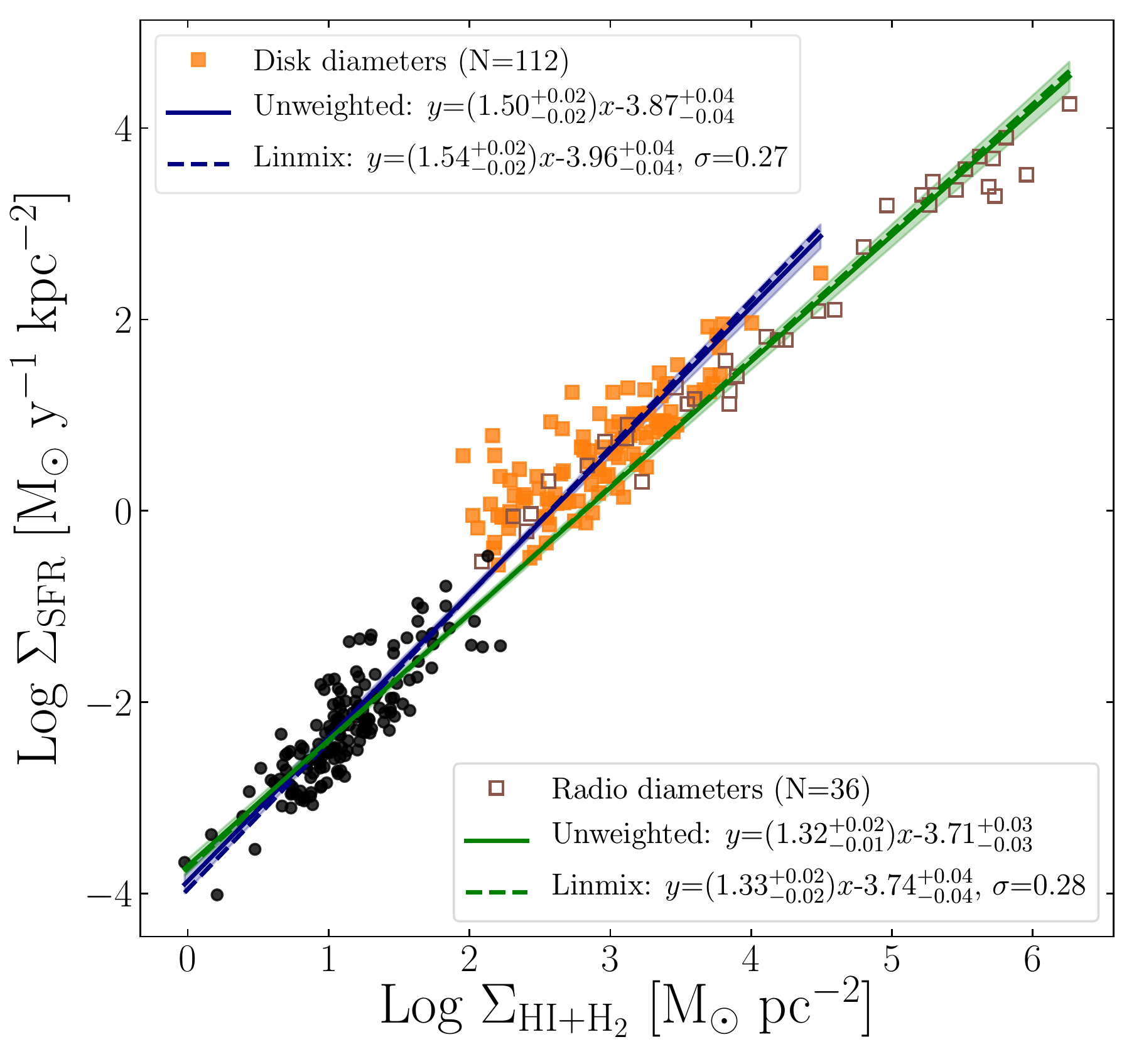}
    \caption{Illustration of the impact of adopting starburst disk diameters measured in the radio at 1.4\,GHz \citep[empty points and green linear fits, from][]{Liu2015} in place of actual disk diameters measured in CO or the infrared dust emission (solid points and blue linear fits).  The radio sizes are only marginally correlated with the CO or infrared sizes, and on average are more than an order of magnitude smaller.  This causes the gas and SFR surface densities to be overestimated by $\sim$2 orders of magnitude on average, producing a population of objects with unphysically high surface densities.  The net effect is to decrease the measured slope of the composite Schmidt law.} 
    \label{fig:diameters_radio}
\end{figure}

Often most of the observed CO and IR emission lies well outside of the radio-determined radii, and when that emission is presumed to lie within the latter the SFR and molecular gas surface densities will tend to be overestimated, by as much as two orders of magnitude.  The net result is illustrated in Figure~\ref{fig:diameters_radio}, where we have plotted our estimates of the surface densities of star formation and gas in a sample of LIRGs and ULIRGs based on disk diameters measured directly in CO and/or infrared (solid squares), and overplotted the surface densities for the same objects that would result if we were to use the radio diameters instead (open squares).  

Using radio diameters causes both the SFR and gas surface densities to be overestimated by identical factors, effectively extending the apparent Schmidt relation by $\sim$2 orders of magnitude along a trajectory with slope $n=1$.  
The net effect on the combined spiral--starburst Schmidt law is to decrease the overall slope by $\sim$0.2\,dex, which is nearly enough to explain the shallower power law measured by \citet{Liu2015}.
This change in slope may be further exaggerated by using radio diameters for normal spirals, which also tend to be systematically smaller but only by tens of percent rather than factors of order ten.  These differences in methodology seem to account for most of the difference between our reported results, but we hasten to emphasize that most of the conclusions in \citet{Liu2015} are unaffected by them.

Our work also revisits and confirms some conclusions previously drawn from systematic studies of total molecular masses and star formation rates in galaxies, and the implied molecular depletion times (e.g., \citet{Saintonge2011, Huang2014, Huang2015}).  We discuss these in the following discussion section.

\section{Discussion:  The Case for a Multi-Modal Star Formation Law}
\label{sec:discussion}

Evidence for a bimodal star formation law has existed for more than a decade \citep{Daddi2010,Genzel2010}, though the dependence of the result on assumptions concerning the \xco{} conversion factor have always raised questions about its reality and interpretation.  While this paper reproduces the earlier evidence for bimodality between the Schmidt laws in normal vs starburst galaxies, three new aspects of our results bolster the case for genuine physical multi-modality, and offer tempting clues to the physical underpinnings of these differences:
\begin{enumerate}
    \item Unlike in previous studies, which mainly showed evidence in the shift in the zeropoint of the Schmidt law and not in its slope, our results show pronounced breaks both in the slope (from $n \sim 1.4$ to $n \sim 1.0$) and the zeropoint (or characteristic star formation efficiencies) of the {\bf relations (by $\sim$0.6 -- 0.7 dex).}
    \item Unlike in previous studies, the breaks in slope and zeropoint persist even when identical values of the \xco{} conversion factor are applied to all galaxies, or when using dust surface densities measured independently of CO.  Adopting lower \xco{} values for the starbursts, as is often advocated, only magnifies the apparent bimodality.
    \item Breaks in the Schmidt law are seen regardless of whether the SFR surface densities are plotted against total (atomic plus molecular) gas surface densities or against molecular surface densities alone.  However a break in slope $n$ is only seen in the total gas density relation; for molecular gas the Schmidt relation is roughly linear for all but the very lowest surface density systems (which fall into a threshold regime, see \citetalias{delosReyes2019}).
\end{enumerate}

It is the combination of the strong character of the breaks in the relation (nearly 0.5 dex in $n$ and a factor of five or larger in zeropoint) and the relative behaviors of the relations in \hi{} $+$ H$_2$ and H$_2$ alone that strengthens our confidence that the results cannot arise from observational errors, sample selection effects, or other spurious sources.  The results also suggest a fairly straightforward physical interpretation.  The shift in slope of the relation is naturally explained as a transition from an ISM in normal disks roughly equally divided between atomic and molecular gas, to a nearly totally molecular-dominated ISM in the starbursts.  This interpretation was introduced by \citet{Bigiel2008} and \citet{Leroy2008} to explain their spatially resolved measurements of the Schmidt law in the SINGS sample, and it provides a natural explanation for our results as well.  The fact that the molecular Schmidt law shows no break in slope is fully consistent with this interpretation.

As discussed previously by many authors, the break in the zeropoint of the relation is less straightforward to interpret, but if taken literally it implies that the characteristic SFR per unit gas mass is several times higher in the starbursts (or rephrased, the characteristic gas depletion times are several times shorter in the starbursts).  This difference applies whether one considers total (atomic plus molecular) or only molecular gas depletion times.  The break in molecular Schmidt laws and depletion times is especially noteworthy.  Numerous studies have suggested that among normal star-forming spirals, such as those studied in \citetalias{delosReyes2019}, the molecular depletion times are roughly constant \citep[e.g.,][and references therein]{Leroy2013}.  Likewise the results in this paper show that starbursts follow a roughly linear (molecular) Schmidt law, and thus also exhibit a roughly constant depletion time, just 5--8 times shorter than in the non-starbursting galaxies.  This suggests that whatever physics is driving the higher efficiencies in the starbursts must be something other than (or in addition to) the transition to a molecular-dominated phase.  A detailed discussion of possible mechanisms lies outside of the scope of this paper, but explanations could include real changes in star formation efficiencies combined with changes in parameters such as the stellar initial mass function (IMF).  For example a relatively top-heavy IMF in the starbursts would produce higher SFRs and infrared luminosities for a given gas mass, because the luminosity is heavily weighted to the most massive stars while most of the stellar mass is contributed by low-mass stars \citep[][and references therein]{Bastian2010}.  It is worth noting however that the five-fold or greater increase in luminosity per unit gas mass would require a dramatic change to the IMF, such as a shallower IMF slope of order unity \citep{Clauwens2016}, an increase in the upper stellar mass limit to $\ge$200\,M$_{\odot}$, or removing all stars with masses below 1\,M$_{\odot}$ \citep{Chabrier2003}.  Such dramatic changes in the IMF seem unlikely but may not be ruled out entirely by observations.

Our analysis leaves some important questions about the star formation law unanswered, and for us one stands out in our minds.  Is the Schmidt law truly bimodal, both in slope and zeropoint, as seems to be strongly suggested by the results in this paper?  Or could the relation vary more continuously across the full range of star formation environments found in galaxies, with the results for nearby normal disks and infrared-luminous and ultraluminous galaxies analyzed here merely representing the two extremes of this variability?  

Part of the motivation for asking these questions is the realization that the two subsets of galaxies in this paper were selected in completely different ways.  The spiral galaxy sample (\citetalias{delosReyes2019}) was based entirely on the availability of spatially-resolved \hi{}, CO, UV, and mid-infrared meaurements.  Although it is by no means a complete sample in a volume-limited sense, it should be a reasonably representative sample of normal star-forming disk galaxies in the local universe.  On the other hand, the starburst sample is dominated by luminous and ultraluminous infrared galaxies, defined as those with total infrared luminosities above $10^{11}$ and $10^{12}$ L$_{\odot}$, or SFRs above 16 and 160 M$_{\odot}$\,yr$^{-1}$ for the calibration used in this paper, i.e., they represent SFR-limited samples.  Most are the products of massive tidal interactions or major mergers, with SFRs well above the typical maximum SFR of $\sim$10\,M$_{\odot}$\,yr$^{-1}$ found in present-day undisturbed galaxies \citep[e.g.,][]{Bothwell2011}.  The starburst galaxies were selected independently of their molecular gas masses, apart from needing to possess reliable CO maps and/or flux measurements.  To what extent could this SFR-biased selection for the starbursts be responsible for what appears as a bimodal star formation law?

Infrared luminosity could manifest itself in more than one way.  Presumably all starbursts go through a cycle in which gas gradually accumulates in the circumnuclear regions, eventually triggering a burst of star formation, but not necessarily in lockstep with the gas supply at the moment.  Either a lag in the triggering of the star formation or a lag in the infrared-luminous phase relative to the instantaneous SFR could lead to an enhancement of the IR luminosity per unit gas mass relative to the intrinsic, time-averaged Schmidt relation.  In principle if this lag were long enough compared to the gas depletion times, the measured infrared luminosities could systematically overestimate the SFR and star formation efficiencies.  We used Starburst99 \citep{Leitherer99} models to investigate the likely magnitude of this effect, but found that the lag times for a significant overestimation of the SFR is too short ($\ll$10 Myr) to be very important.  

\begin{figure*}
    \centering
    \epsscale{1.1}
    \plottwo{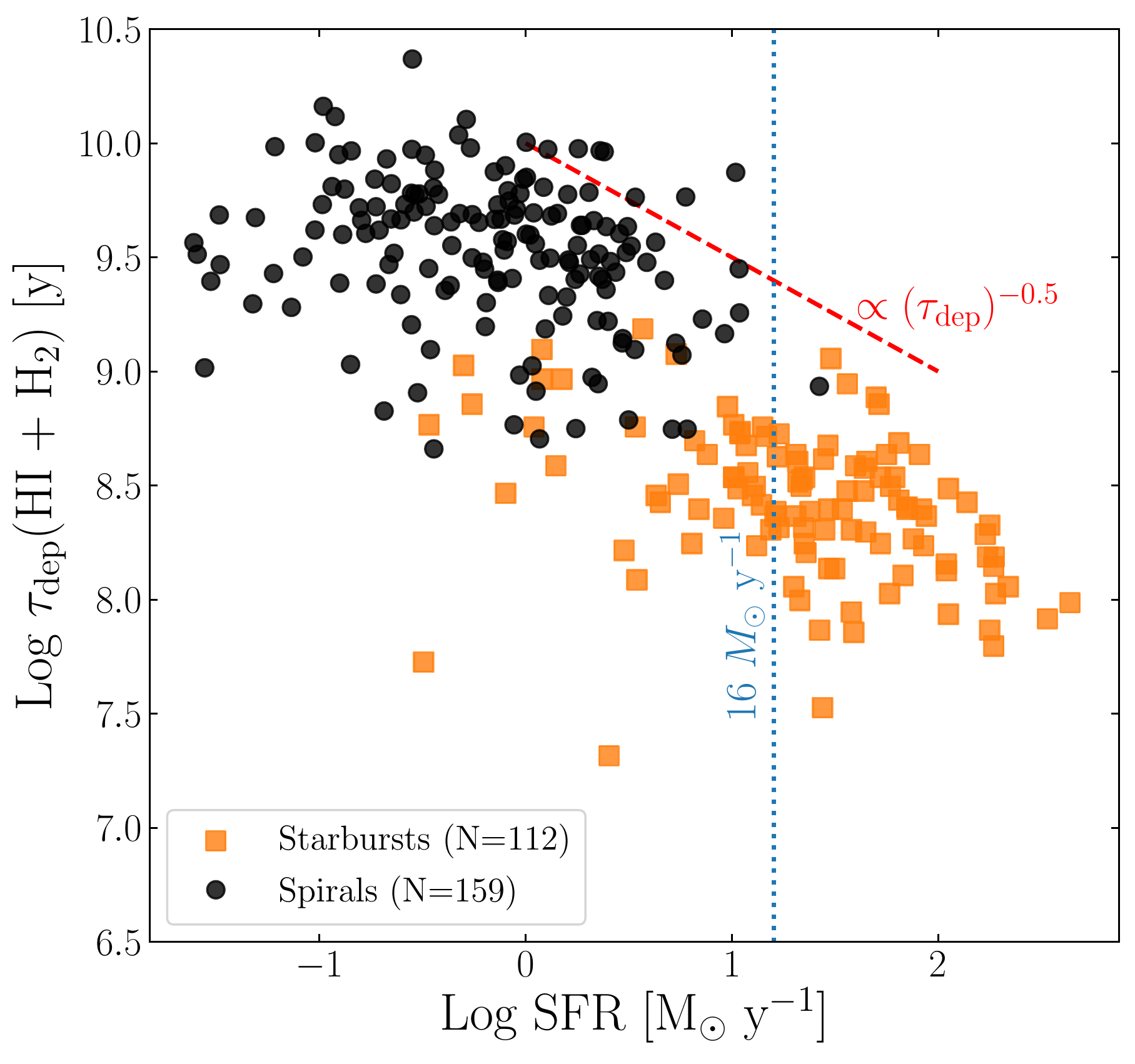}{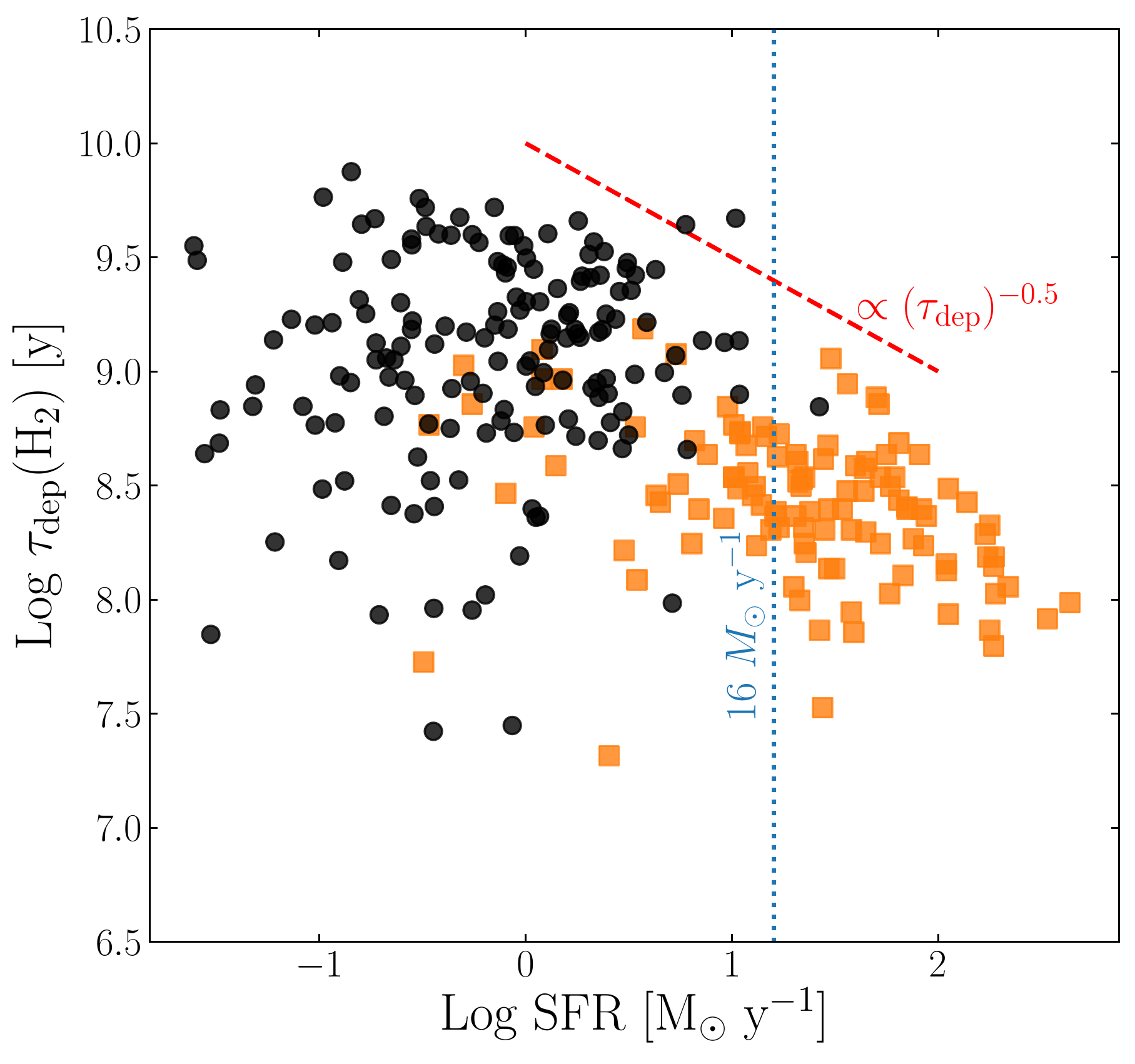}
    \caption{Left:  Total gas depletion times for normal spiral galaxies (solid black points) and starbursts (orange squares) as a function of the current SFR.  Right:  Similar but showing the molecular gas depletion times.  A Milky Way X(CO) factor was applied uniformly to both samples for this comparison. For illustrative purposes, the red dashed line indicates a slope of $-0.5$, and the vertical blue dotted line denotes $\mathrm{SFR}=16~M_{\odot}~\mathrm{y}^{-1}$ (corresponding to the minimum luminosity of a LIRG).}
    \label{fig:depletion_vs_sfr}
\end{figure*}

Another, empirical way to check for a possible bias is to adopt the approach of \citet{Saintonge2011} and plot the gas depletion times for the galaxies in our combined sample as a function of the absolute SFR, as shown in Figure~\ref{fig:depletion_vs_sfr}.  As expected, the starbursts as a class show systematically lower depletion times (consistent with the definition of a starburst).  If the selection of LIRGs and ULIRGs were adding a further selection bias lowering the depletion times, we might expect a sharp drop in timescales at a SFR of 16\,M$_{\odot}$\,yr$^{-1}$ (the SFR corresponding to the minimum luminosity of a LIRG), denoted by the dashed vertical line in the panels.  We observe no such transition; instead the difference in depletion times appears to be much more associated with the identification of the regions as starbursts than with any threshold in luminosity selection.  We do find it interesting that the distribution of depletion times appears to have an upper envelope, extending from the normal spirals to the LIRGs and ULIRGs, with a slope very close to $-$0.5, corresponding to $\mathrm{SFR}\propto M{_{\mathrm{gas}}}^2$.  The physical significance of this upper limit (or absence of galaxies with longer depletion times) is unclear.  

Although we have failed to identify clear evidence for the possibility that the bimodality seen in our results \citep[and the previous results of][]{Daddi2010,Genzel2010} arises from sample selection effects, there are other reasons to consider the possibility that the changes in the form of the Schmidt law seen between normal spiral galaxies and LIRGs/ULIRGs may be more continuous than discrete.  Perhaps the strongest hints come from large-scale surveys of the molecular gas in both local and distant galaxies \citep[for a review, see][]{Tacconi2020}.  Although these measurements lack the spatial resolution needed to analyze surface densities, they provide very accurate measurements of disk-averaged molecular depletion times, which for a linear star formation law are insensitive to spatial sampling.  The COLD GASS study \citep{Saintonge2011} found that the molecular depletion times across a wide range of galaxy types vary continuously by nearly two orders of magnitude, as a function of SFR per unit stellar mass (specific star formation rate).  Local spirals and LIRGs/ULIRGs occupy the high and low extremes in the full range of depletion times, but galaxies in the sample fill the full range.  This work was followed up by \citet{Huang2014, Huang2015}, who found that the correlations between molecular depletion times and SFR are present in both integrated and spatially-resolved measurements, with significant differences as functions of stellar surface density, between bulge and disk regions, and with spiral arm and bar morphologies.  These studies differ from the present investigation in terms of the measureables (i.e., total molecular gas masses instead of surface densities), but because of the near-linearity of the molecular Schmidt law in all galaxies any trends in H$_2$ relations and depletion times alone should be similar in the two cases.

It is not currently possible to perform the analogous test for the Schmidt surface density law, because samples of galaxies with spatially resolved \hi{} and CO measurements are not available for a sufficiently diverse parent sample.  The results presented in \S\ref{sec:dustSchmidt} using dust surface densities as a proxy for gas, however, suggest that a follow-up study extending our analysis to a much broader and more diverse set of galaxies is feasible.  We are undertaking such an investigation for the final paper in this series.

\section{Summary}
\label{sec:conclusion}

In these two papers, we have revisited the widely cited \citetalias{Kennicutt1998} study of the disk-averaged star formation law in galaxies, extending from normal non-starbursting disk and dwarf galaxies (\citetalias{delosReyes2019}) to the most extreme infrared-luminous starburst galaxies and other circumnuclear starbursts (this paper).  The new study has taken advantage of a roughly threefold increase in the galaxy sample, vastly improved measurements of dust-corrected SFRs, and a more diverse sample of galaxies, especially in \citetalias{delosReyes2019}.  

A key conclusion from \citetalias{Kennicutt1998} was that the relation between SFR and total (atomic plus molecular) gas surface densities for a combined sample of normal and starburst galaxies was well represented by a single non-linear Schmidt power law.  This result is confirmed here, with the slope of the best-fitting Schmidt law now being $n = 1.50 \pm 0.05$ as opposed to $n = 1.4\pm0.1$ in \citetalias{Kennicutt1998} and $n \sim 1.2$ by \citet{Liu2015}.  A single power-law approximation remains as a credible star formation ``recipe" for models and numerical simulations.

The new data reveal considerable deeper complexity in the Schmidt law, however.  When the relations for normal and starburst galaxies are fitted separately there is evidence for strong breaks both in the slopes (1.4 to 1.0) and zeropoints ($\sim 0.6$~dex).  Similar breaks in zeropoints are seen in the molecular gas Schmidt law.  These breaks are observed for all plausible choices of values for the \xco{} conversion factors, even ones in which the same conversions are applied to all galaxies, and are also seen when the SFR is measured as a function of the interstellar dust surface densities.  Our results on the bimodality in Schmidt law zeropoionts are broadly consistent with those reported previously by other groups \citep[e.g.,][]{Daddi2010,Genzel2010}. Our results are very difficult to reconcile with the adoption of very low \xco{} conversions for the infrared luminous starburst regions often suggested by modelling of the CO rotational spectra of starburst galaxies.

Comparisons of the star formation rate vs gas density relations for subsets of the starburst galaxies selected in different ways show small (if any) systematic differences, and comparisons of the methods used to measure the SFRs also show insufficient differences to explain the large changes in slope and zeropoint of the Schmidt laws between the normal and starburst galaxy samples.  Thus we tentatively conclude that the differences are due to a physical origin. The change in slope of the law is most likely associated with the transition from a mixed atomic and molecular cold ISM in normal disks, transitioning to a nearly fully molecular ISM in the starbursts.  The break in zeropoints or gas depletion times (also seen in the molecular depletion times) must have another explanation.  Possibilities include a genuine change in local star formation efficiencies and/or a top-heavy stellar IMF.  We also consider the possibility that the luminosity-based selection of many of our starburst galaxies might contribute to an apparent bimodality in the star formation law, but we are unable to identify any clear observational evidence for the presence of a large bias of this type.

We demonstrate that the properties of the measured star formation law are sensitive to the methods used to define the diameters of the actively star-forming disks, and this sensitivity should be considered in future studies when comparing different sets of results in the literature.

An important question left unresolved by this study is whether the differences in Schmidt laws seen between normal non-starbursting disk galaxies and the infrared-luminous circumnuclear starbursts reflect a true bimodality, or merely the extremes of a more continuous underlying change in the relation.  Answering this question requires observations of galaxies with properties intermediate between those studied here and in \citetalias{delosReyes2019}.  We intend to pursue this issue in Paper III of this series.

\acknowledgments

The authors wish to thank an anonymous referee for a report which led to significant improvements in this paper. 
Early phases of this research were supported in part by the STFC through a consolidated grant to the Institute of Astronomy, University of Cambridge.
M. A. de los Reyes also acknowledges the financial support of the NSF Graduate Research Fellowship Program.
Finally, we would like to express our deep gratitude to the staff at academic and telescope facilities, particularly those whose communities are excluded from the academic system, but whose labor maintains spaces for scientific inquiry.

\software{
Matplotlib \citep{matplotlib}, 
Linmix \citep{Kelly2007}, 
Astropy \citep{astropy}
}

\appendix
\section{Gas surface density references}
\label{appendix:gasrefs}

Here we list the references from which CO masses for starburst galaxies (Table~\ref{tab:starburst}) were obtained.

1: \citet{Ueda2014};
2: \citet{Downes1998};
3: \citet{Garcia-Burillo2012};
4: \citet{Sanders1991};
5: \citet{Scoville1985};
6: \citet{Bryant1999};
7: \citet{Gao1999};
8: \citet{Jogee2005};
9: \citet{Yamashita2017};
10: \citet{Horellou1997};
11: \citet{Albrecht2007};
12: \citet{Mirabel1990};
13: \citet{Gracia-Carpio2008};
14: \citet{Kohno2003};
15: \citet{Sakamoto2007};
16: \citet{Sakamoto1999};
17: \citet{Aalto1994};
18: \citet{Salak2016};
19: \citet{Lo1987};
20: \citet{Nakai1987};
21: \citet{Planesas1991};
22: \citet{Gracia-Carpio2007};
23: \citet{Sakamoto2014};
24: \citet{Kenney1992};
25: \citet{Horellou1995};
26: \citet{Wiklind1995};
27: \citet{Sargent1991};
28: \citet{Momose2010};
29: \citet{Lundgren2004};
30: \citet{Evans2002};
31: \citet{Lo1997};
32: \citet{Chung2009};
33: \citet{Scoville1997};
34: \citet{Gao2004};
35: \citet{Aalto1995};
36: \citet{Pan2013};
37: \citet{Sanders1988}

\section{Statistical fitting techniques}
\label{appendix:stats}
In this section, we consider the validity of the linear regression models used in this paper.
Before discussing each regression technique, we summarize the aspects of our data relevant to linear regression.
First, although the star formation law is empirical, there is some physical motivation for a linear relationship between $\log\Sigma_{\mathrm{gas}}$ and $\log\Sigma_{\mathrm{SFR}}$ in which $\Sigma_{\mathrm{SFR}}$ is physically expected to be the dependent ``response'' variable (see \S\ref{sec:intro}).

There are measurement errors in both the $x$ and $y$ quantities.
As discussed in Section~\ref{sec:uncertainties}, we estimate the measurement uncertainties on both the \emph{absolute} gas masses and SFRs for the starburst and spiral galaxies to be $\sim0.1$~dex.
However, the gas and SFR \emph{surface densities} for the starburst galaxies are significantly more uncertain due to additional uncertainty of up to $\sim0.08$~dex in the measured diameters.
Since both $\Sigma_{\mathrm{gas}}$ and $\Sigma_{\mathrm{SFR}}$ depend on diameter in the same way, the $x$ and $y$ errors are directly correlated.

Finally, star formation is a complex phenomenon, and gas density may not be the only factor driving star formation rate on galactic scales. 
We might therefore expect some intrinsic dispersion in $\Sigma_{\mathrm{SFR}}$ at a given $\Sigma_{\mathrm{gas}}$ that cannot be explained by measurement uncertainty alone.
The underlying distribution of this intrinsic scatter is unknown.

\subsection{Unweighted least squares}
We begin by considering the simplest linear regression method: unweighted ordinary least squares (OLS).
Unweighted OLS minimizes the mean squared $y$-residuals and weights all points equally, without accounting for any statistical uncertainties. 
Since the measurement errors in $\Sigma_{\textrm{gas}}$ and in $\Sigma_{\textrm{SFR}}$ are of roughly the same order of magnitude, this method provides a reasonable first-order approximation to compare with other fits.

\begin{figure*}[t!]
    \centering
    \epsscale{1.1}
    \plottwo{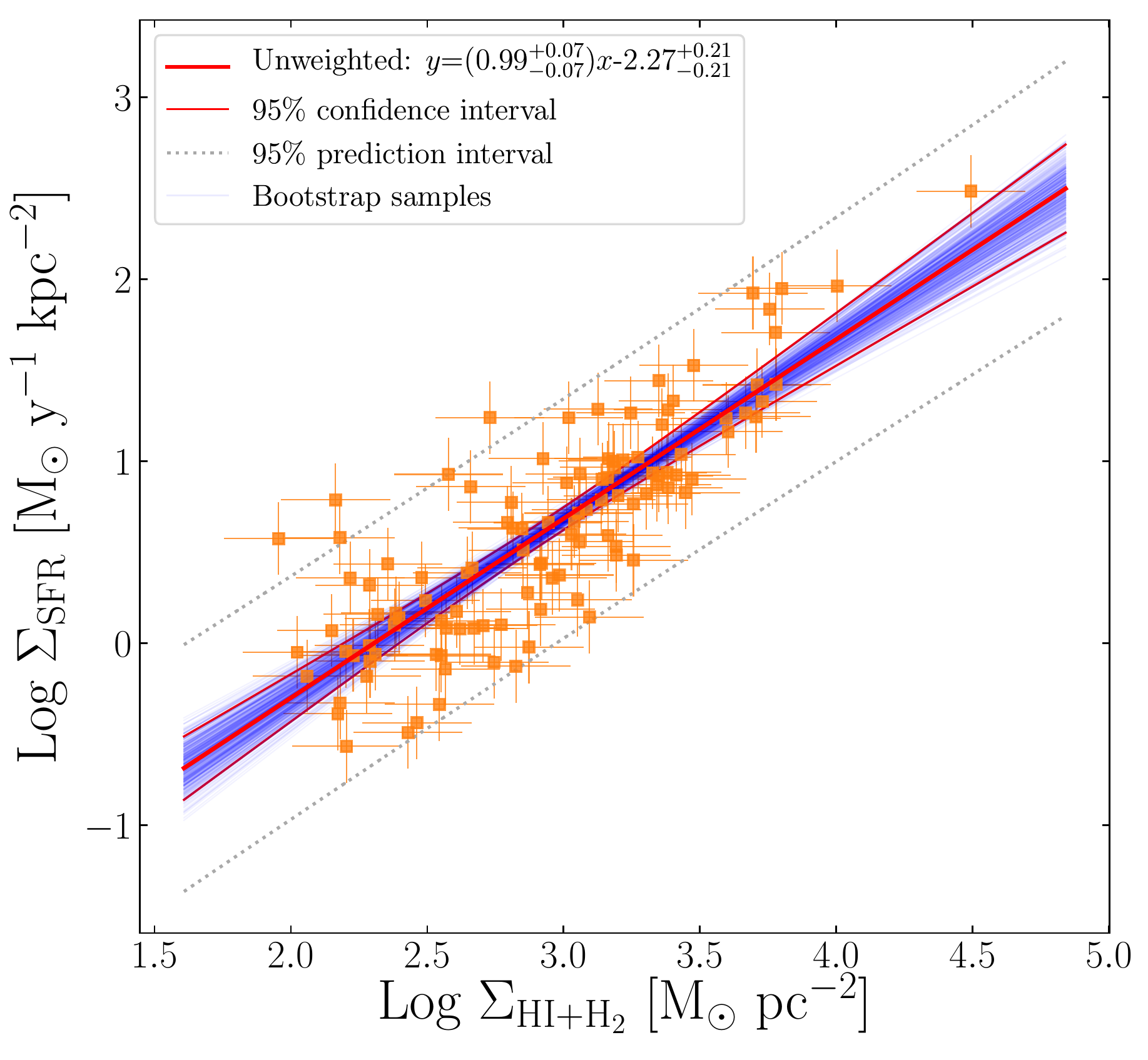}{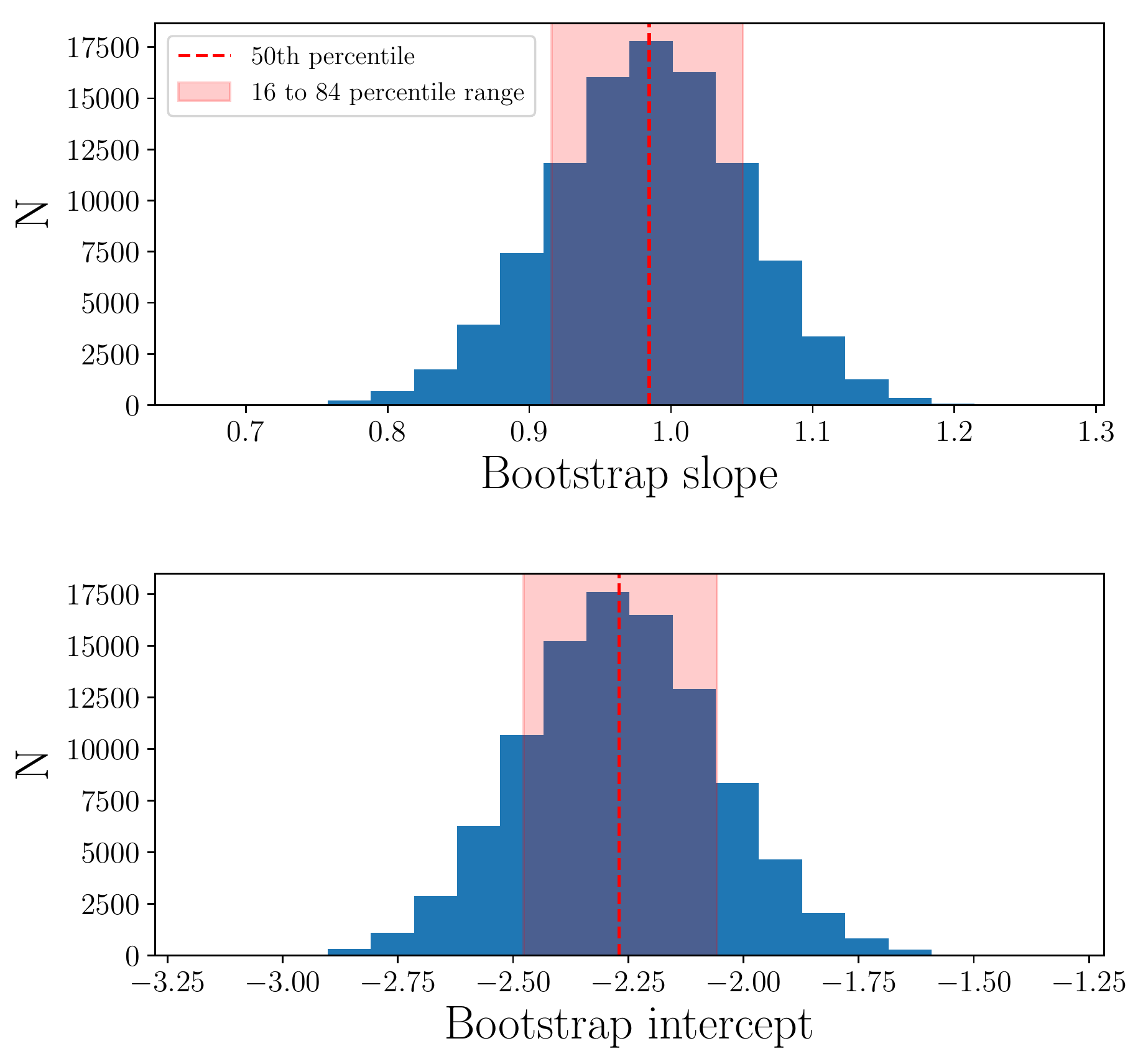}
    \caption{Left: Unweighted OLS for the starburst sample (similar to Fig~\ref{fig:starburst_sample_only}). Thin blue lines indicate fits drawn from the bootstrap distribution, while the thick red solid line indicates the bootstrap median fit. The errors reported in the parameters are the 16--84th percentile values from the bootstrap distributions. The thin red lines and gray dotted lines respectively indicate the 95\% confidence and prediction bands about the non-bootstrapped fit, both computed under the assumption that the regression parameters are normally distributed. Right: The bootstrap distributions of the slope (top) and intercept (bottom) parameters.}
    \label{fig:unweightedtest}
\end{figure*}

Given our relatively small sample size, it is possible that some outliers may have disproportionately high leverage on the resulting fit.
We test this using non-parametric paired bootstrap resampling, in which we create $100000$ bootstrap samples by drawing randomly with replacement from the original set of points.
We then fit each bootstrap sample using unweighted OLS.
Figure~\ref{fig:unweightedtest} shows the results of this test, using the star formation law for our sample of starburst galaxies (as in Fig~\ref{fig:starburst_sample_only}). 
In the left panel, we plot $400$ fits drawn from the bootstrap distribution as thin black lines, with the median bootstrap fit shown as a thick red line.
For illustration, we also indicate the ``normal-theory'' 95\% confidence and prediction bands \citep[the 95\% confidence and prediction intervals assuming that the regression parameters are normally distributed; e.g.,][]{Fox2002} about the median fit.
The right panel shows the bootstrap distributions of the slope and intercept parameters, both of which are approximately normal.
Throughout this work, all parameters from unweighted OLS fitting are reported as the median bootstrap values with uncertainties denoting the 16th to 84th percentile range.

%

Beyond checking for points with disproportionate leverage, bootstrapping also allows us to estimate errors in the regression parameters without directly using our estimates of the measurement uncertainties or making any assumptions about the underlying error distributions.
Bootstrapping does rely heavily on other assumptions about the model---in particular, it assumes that the original sample is reasonably representative of some ``true'' population, which is difficult to fully verify for any astrophysical population.
It also assumes that the underlying linear model is accurate, which does not account for any potential intrinsic scatter.
However, the bootstrapped unweighted OLS method does address many of the aspects of our data while making relatively few assumptions.

\subsection{Linmix}
On the other end of the spectrum from unweighted OLS and its minimal assumptions, linear regression that includes intrinsic dispersion in addition to both $x$- and $y$-errors is a fundamentally under-constrained problem that requires a number of constraining assumptions to solve.
The ``linmix'' estimator \citep{Kelly2007} is one attempt to address this.
Linmix is a maximum likelihood estimator, in which the likelihood function is computed by convolving multiple ``hierarchical'' Gaussian distributions.

In this model, any measured data point $(x,y)$ can be drawn from a 2D Gaussian distribution $P_{1}$ with some ``true'' mean $(\xi,\eta)$ and covariance matrix determined from measurement uncertainties.
The ``true'' value of the dependent variable $\eta$ can be drawn from a Gaussian distribution $P_{2}$ with mean $\beta\xi + \alpha$ and variance $\sigma^{2}$, where $\beta$ and $\alpha$ are the regression coefficients, and $\sigma^{2}$ is the intrinsic dispersion.
The ``true'' value of the independent variable $\xi$ is assumed to be drawn from a weighted sum of $K$ Gaussian distributions $P_{3}$, since a large enough number of Gaussian functions can approximate any true distribution.
(As in \citetalias{delosReyes2019}, we use $K=2$ Gaussian distributions in this work; the addition of more distributions has a negligible effect on our results.)
The distributions $\{P_{1},P_{2},P_{3}\}$ are convolved hierarchically to obtain the full likelihood.
We assume that the prior distributions for the model parameters $\{\beta,\alpha,\sigma^{2}\}$ are uniform, then run a Markov chain Monte Carlo (MCMC) algorithm to sample from the posterior distributions until convergence is reached.

\begin{figure}[h!]
    \centering
    \epsscale{0.7}
    \plotone{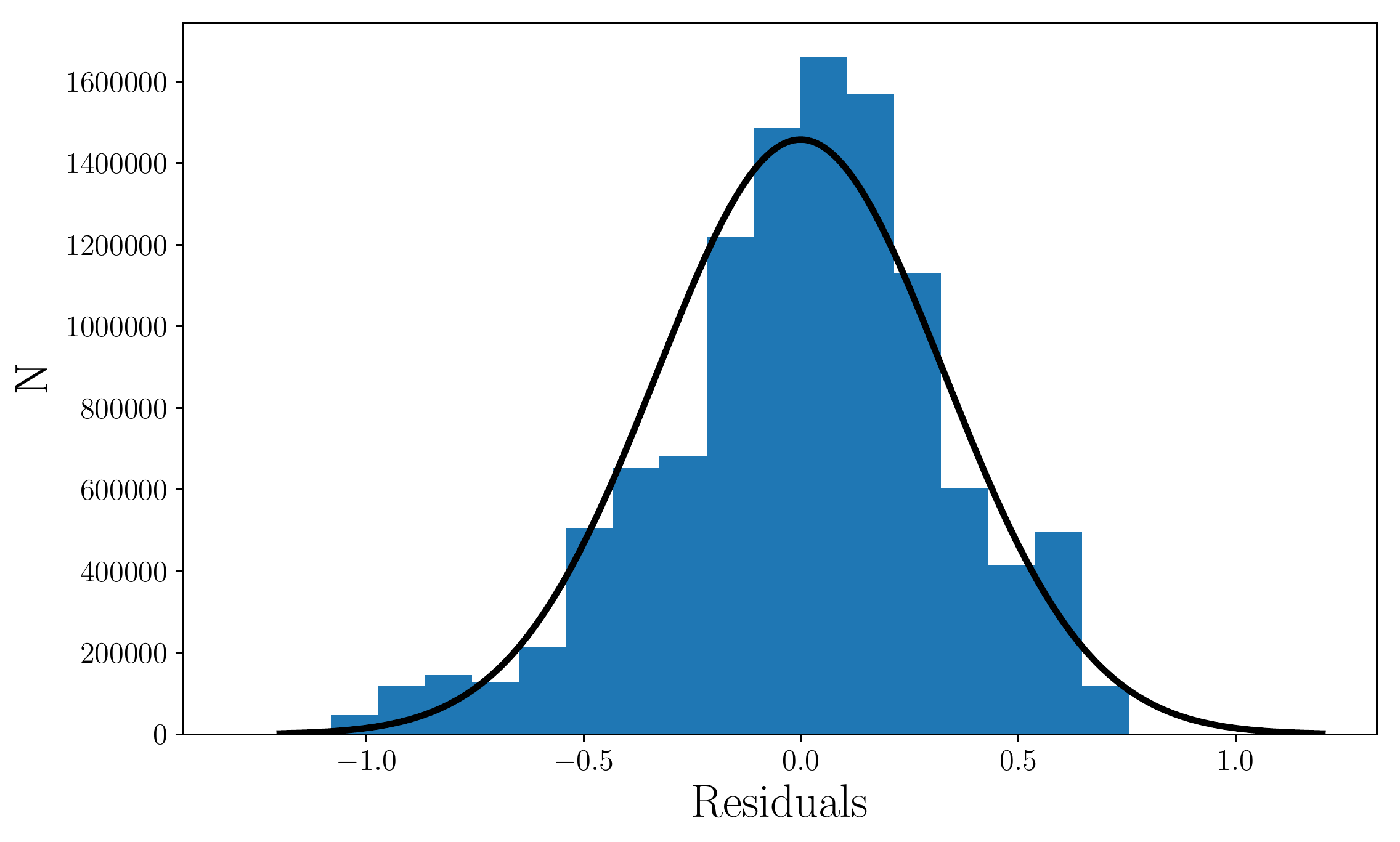}
    \caption{Distribution of the residuals about the bootstrap unweighted fits shown in Fig~\ref{fig:unweightedtest}. The best-fit normal distribution (black line) is shown for comparison.}
    \label{fig:checkscatter}
\end{figure}

The linmix model makes a number of assumptions.
First, linmix assumes that the intrinsic scatter about the linear relation is normal in the $y$-direction.
We check this by plotting the distribution of $y$-residuals about the bootstrap unweighted fits in Figure~\ref{fig:checkscatter}, which shows that while the scatter is not exactly Gaussian, it is reasonably well-approximated by a normal distribution.

Unlike bootstrapped unweighted OLS, the linmix method explicitly relies on assumptions about the measurement uncertainties.
It assumes that the measurement errors in the $x$- and $y$-directions are normally distributed---while the error distributions are unlikely to be perfectly normal, particularly in log space, a full statistical analysis of non-normal error distributions is beyond the scope of this paper.
Although \citet{Kelly2007} note that linmix may possibly account for correlated errors (by assuming that points are drawn from a 2D Gaussian distribution with a full covariance matrix), its performance on datasets with correlated errors has not been verified.

\begin{figure*}[t!]
    \centering
    \includegraphics[width=0.49\textwidth]{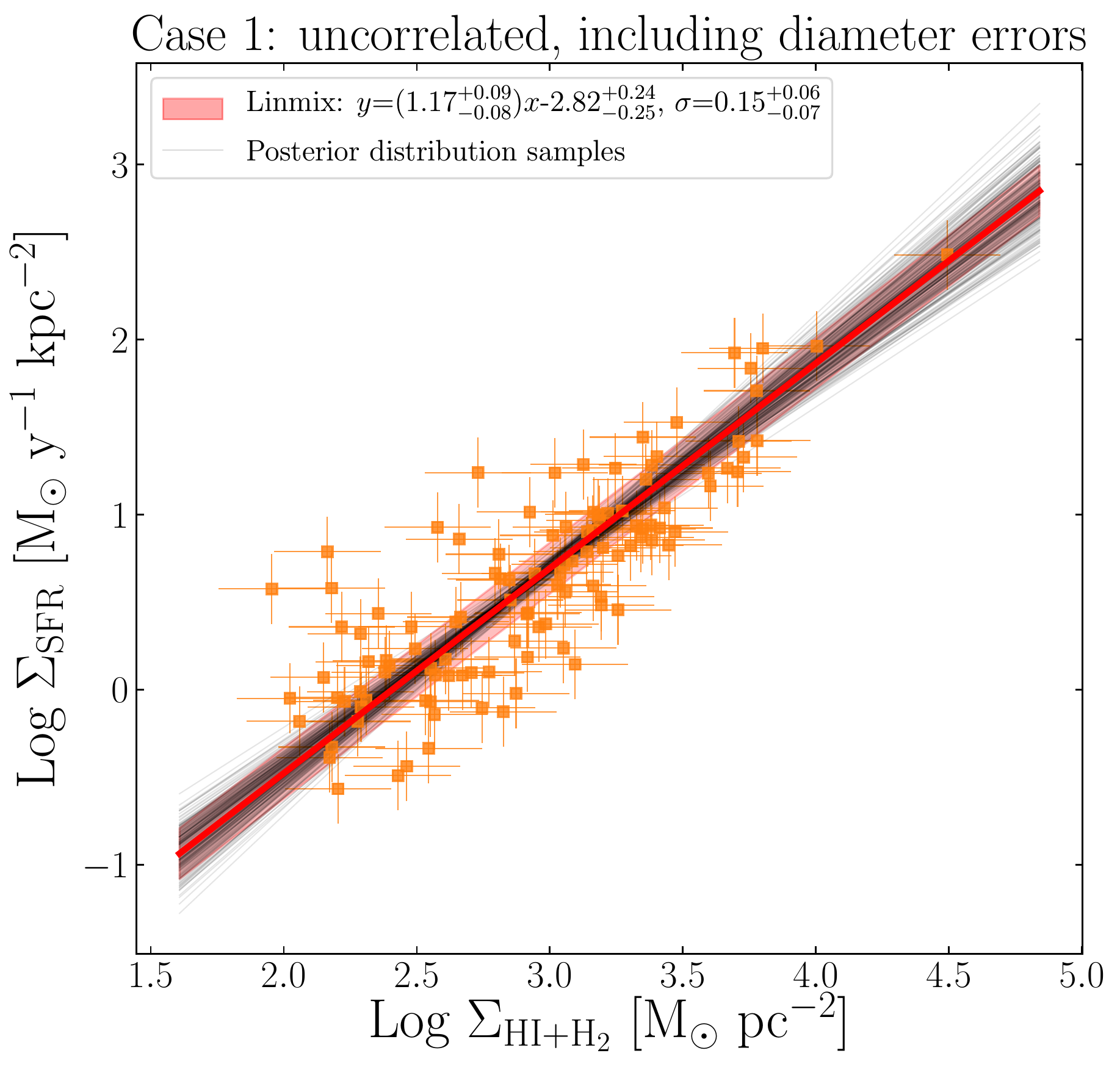}
    \includegraphics[width=0.49\textwidth]{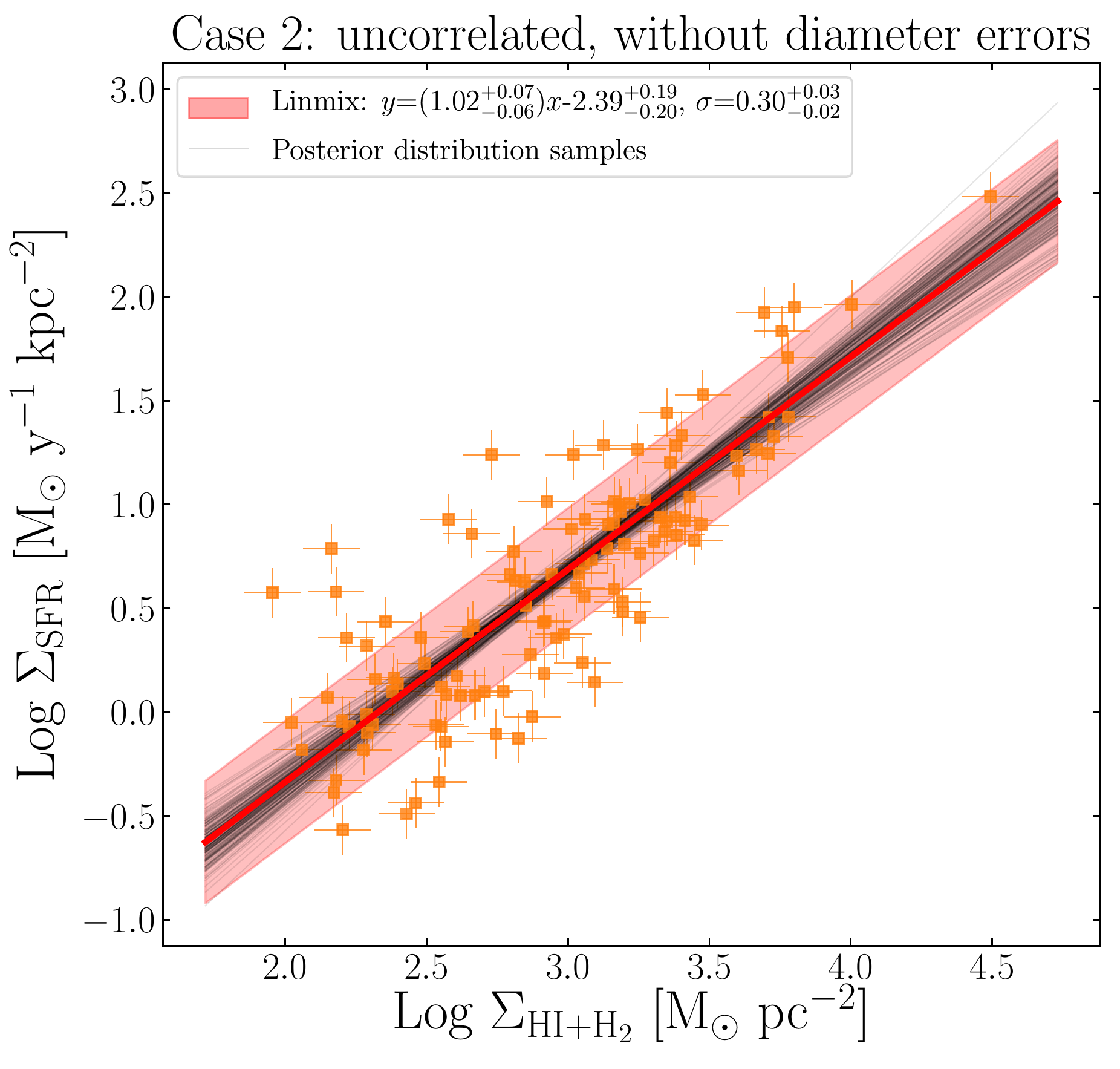}
    \includegraphics[width=0.49\textwidth]{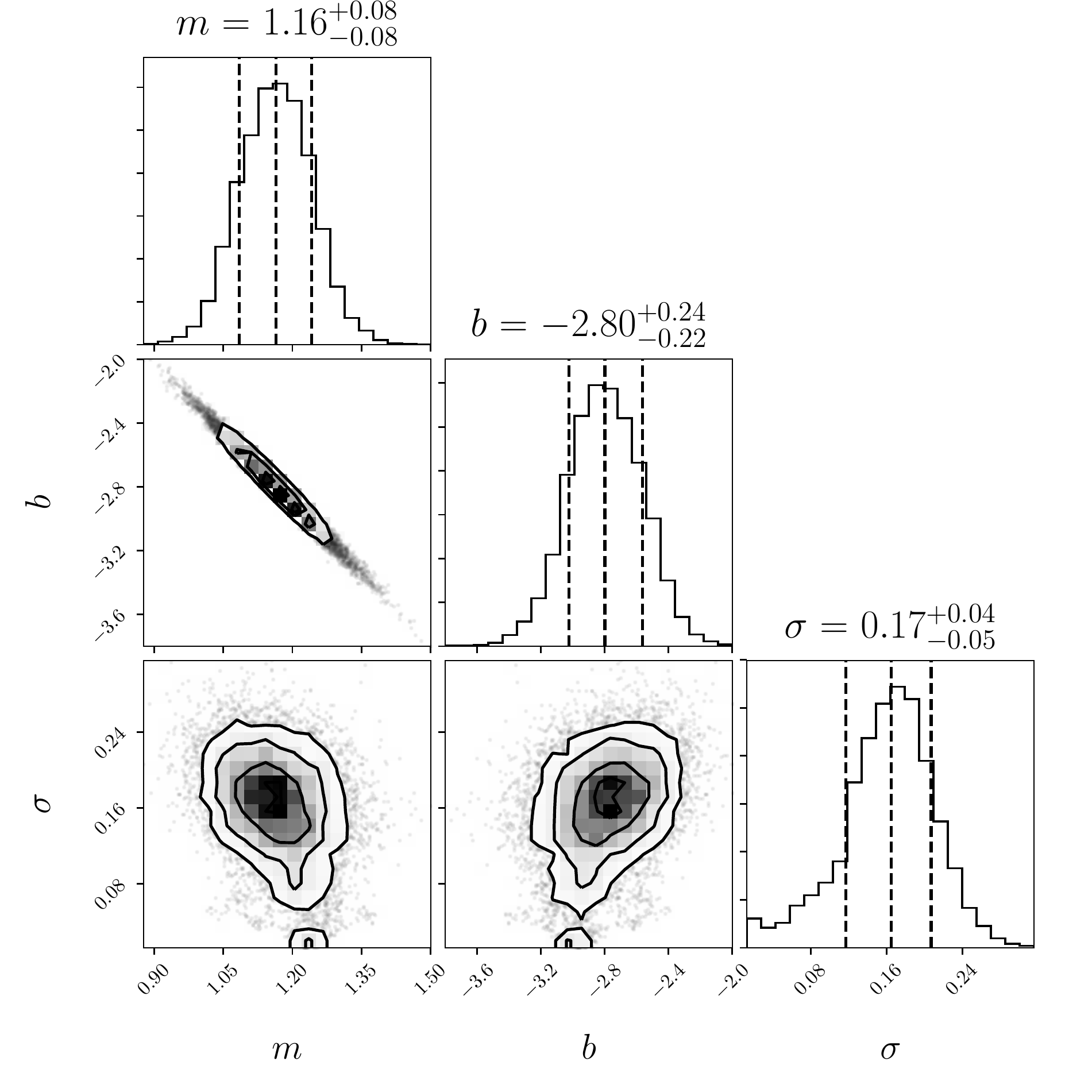}
    \includegraphics[width=0.49\textwidth]{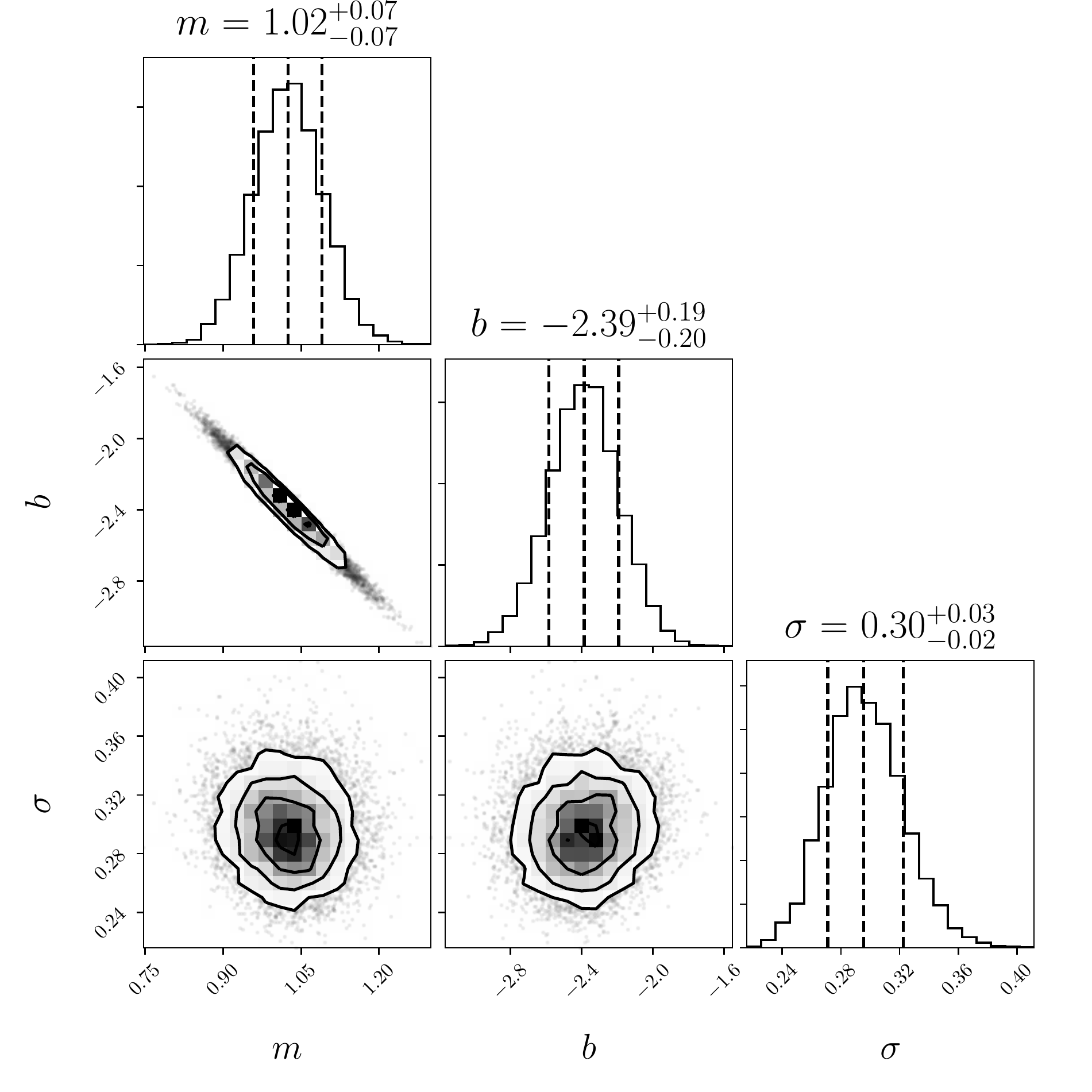}
    \caption{Top left: Linmix fit for the starburst sample (similar to Fig~\ref{fig:starburst_sample_only}), assuming that $x$ and $y$ errors are uncorrelated. The thick red line indicates the median fit from the posterior distribution, while the red shaded region indicates the median estimate of the intrinsic scatter $\sigma$.  
    The thin black lines show fits drawn from the posterior distributions.
    Bottom left: Corner plot showing the posterior distributions of the regression parameters. The 16--84th percentile ranges are marked with black dashed lines.
    Right panels: Same, but assuming no diameter errors.}
    \label{fig:linmixtest}
\end{figure*}

Since the measurement errors in $\Sigma_{\mathrm{gas}}$ and $\Sigma_{\mathrm{SFR}}$ are correlated in the same direction as the intrinsic correlation between gas mass and SFR, treating the errors as uncorrelated will overestimate the slope \citep[see, e.g., Equation 8 of][]{Kelly2007}.
We therefore test the linmix algorithm in two extreme cases: first, including the errors in the diameters but assuming the $x$ and $y$ errors are uncorrelated; second, assuming no error in the diameters at all.
The first case will provide an upper limit for the slope, while the second case will test linmix's sensitivity to the magnitude of the errors.
We test linmix by fitting the star formation law for starburst galaxies in Figure~\ref{fig:linmixtest}.
The left panels show the result for the case with full uncorrelated errors, while the right panels show the result for the case with smaller errors (i.e., excluding the errors in the diameters).
The bottom panels illustrate the posterior distributions of the regression parameters as corner figures, while the top panels show the median fit along with fits drawn from the posterior distributions.

In both cases, linmix appears to provide reasonable fits to our dataset. 
Assuming that the errors are uncorrelated (left panels of Figure~\ref{fig:linmixtest}) yields a higher slope and lower $y$-intercept than the bootstrapped unweighted fits, as expected.
This sets an upper limit of $1.17$ on the slope of the star formation law for the starburst galaxies.
On the other hand, the second case, which assumes smaller errors by setting the errors on the diameter to zero (right panels of Figure~\ref{fig:linmixtest}), yields regression parameters that are inconsistent with the first case and instead agree (within parameter uncertainties) with the bootstrapped unweighted fits.
The estimated intrinsic scatter varies significantly between the two cases: $\sigma=0.15$~dex in the first case, compared to $\sigma=0.30$~dex in the second. 

The discrepancy between these two cases suggests that linmix is relatively sensitive to our knowledge of the measurement uncertainties.
For the spiral galaxies in \citetalias{delosReyes2019}, our main sources of statistical measurement uncertainty were reasonably well-understood: we made many of the measurements ourselves, and the diameters of the star-forming regions were well-defined, so the use of the linmix algorithm was more appropriate.
For the starburst galaxies in this work, we use literature values for both IR and CO fluxes, and the uncertainties in the star-forming diameters are much larger.
We believe that the linmix method is not ideal in this case.
Throughout this paper, we plot the median linmix fit for illustration, but we do not consider the linmix estimates the default.

\subsection{``Bivariate'' least squares}
Another frequently-used linear regression technique for fitting data with both $x$- and $y$-errors is ``orthogonal distance regression,'' a variation on least squares regression.
In this model, rather than minimizing the sum of squared $y$-residuals, the regression aims to minimize the sum of the orthogonal distances between each point and the linear fit.
Estimates of the $x$ and $y$ measurement errors can also be used to weight each point when the $x$ and $y$ measurements have unequal precision \citep[see, e.g.,][]{Boggs1992}.
\citetalias{Kennicutt1998} used this method to estimate the regression parameters of the global star formation law.
For consistency with \citetalias{Kennicutt1998}'s terminology, we will also refer to this as ``bivariate'' regression.

\begin{figure*}[h!]
    \centering
    \epsscale{1.1}
    \plottwo{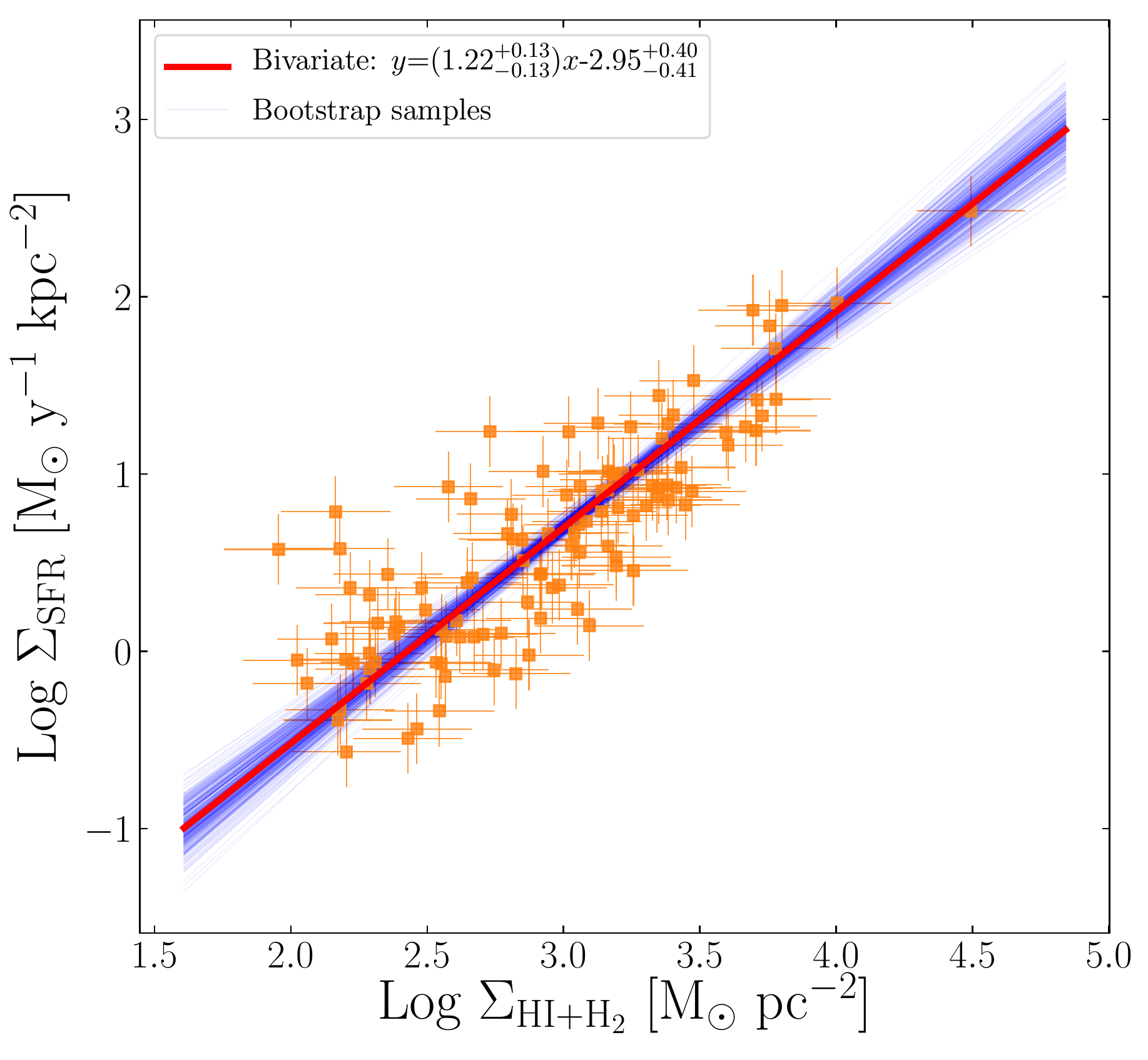}{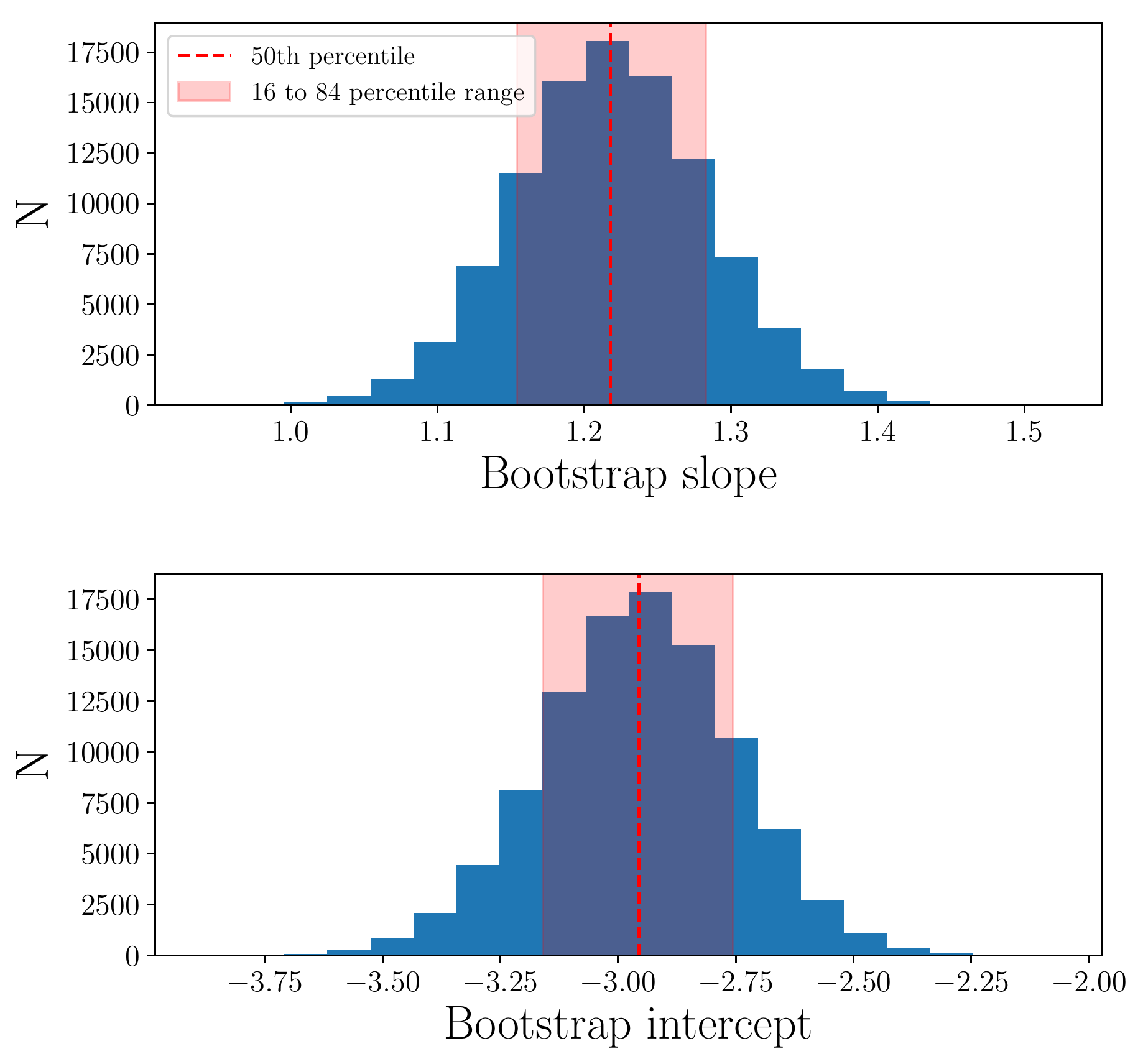}
    \caption{Similar to Fig~\ref{fig:unweightedtest}, but for ``bivariate'' regression. Computing confidence and prediction bands about a bivariate linear fit is nontrivial, and we do not estimate them here.}
    \label{fig:bivariatetest}
\end{figure*}

This model assumes that the $x$- and $y$-errors $\{\delta,\epsilon\}$ are uncorrelated and normally distributed.
It also assumes that the ratio of the standard deviations $d_{i}=\sigma_{\epsilon_{i}}/\sigma_{\delta_{i}}$ is known to a reasonable extent; \citet{Boggs1992} note that even if the values of $d_{i}$ is not known exactly, bivariate regression still provides acceptable parameter estimates and confidence intervals if the $d_{i}$ are known within a factor of 2.
As before, some points may have disproportionately high leverage on the resulting fit, so we test this by creating 100000 bootstrap samples.
Figure~\ref{fig:bivariatetest} shows the results of this test.

Like the linmix estimator, bivariate regression relies on the incorrect assumption that the $x$ and $y$ errors are uncorrelated, likely producing a spurious increase in the observed correlation.
Bivariate regression also assumes that there is no intrinsic dispersion in the linear relation; in the presence of such dispersion, bivariate regression becomes biased and tends to overestimate the slope \citep[e.g.,][]{Akritas96, Carroll96}.
We find that bivariate regression indeed produces a systematically higher slope than unweighted OLS for all the fits in this work. 
This suggests that the star formation law has some intrinsic dispersion, which is expected if factors other than $\Sigma_{\mathrm{gas}}$ help drive $\Sigma_{\mathrm{SFR}}$.
Since bivariate regression appears to produce biased estimates of regression parameters, we do not use this method.
We describe it here primarily for completeness in comparing with \citetalias{Kennicutt1998}. 

\subsection{Summary of linear fitting methods}

We have shown that both linmix and bivariate regression are relatively sensitive to assumptions about measurement uncertainties in our dataset.
This is particularly relevant since the uncertainties in the diameters of the star-forming regions are significant for the starburst galaxies, which means the measurement uncertainties in $\Sigma_{\mathrm{gas}}$ and $\Sigma_{\mathrm{SFR}}$ are strongly correlated.
We therefore treat the regression parameters from bootstrapped unweighted OLS as the default throughout this paper, although for completeness we show both the unweighted and linmix fits in all plots with linear fits.

We note that in \citetalias{delosReyes2019}, when measuring the star formation law for spiral and dwarf galaxies, we reported the results from linmix as the fiducial best-fit parameters.
We suggest the reader be cautious in making direct comparisons between the (linmix) regression parameters discussed in the text of \citetalias{delosReyes2019} and the (bootstrapped unweighted) parameters in this paper.
Throughout this work, when comparing the spiral and starburst samples we will discuss the bootstrapped unweighted OLS results for both samples.
Finally, we note that the reported errors on the regression parameters are likely underestimated regardless of what fitting technique is used, since we cannot fully account for the systematic uncertainties that likely dominate the error budget (see Section~\ref{sec:uncertainties}).

\bibliographystyle{aasjournal}
\bibliography{bibliography}

\end{document}